\documentclass[openacc]{rstransa}%%%%where rstrans is the template name

\usepackage{bm}% bold math
\usepackage{amssymb}

\begin{document}

%%%% Article title to be placed here
\title{A practical method for estimating coupling functions in complex dynamical systems}

\author{%%%% Author details
Isao T. Tokuda$^{1}$,  Zoran Levnajic$^{2}$, 
and Kazuyoshi Ishimura$^{1}$}

%%%%%%%%% Insert author address here
\address{$^{1}$
Department of Mechanical Engineering, Ritsumeikan University, Kusatsu, Japan\\
$^{2}$
Complex Systems and Data Science Lab, Faculty of Information Studies in Novo Mesto, Novo Mesto, Slovenia}

%%%% Subject entries to be placed here %%%%
\subject{Nonlinear dynamics, coupled oscillators, data analysis}

%%%% Keyword entries to be placed here %%%%
\keywords{Coupling function, phase dynamics, parameter estimation}

%%%% Insert corresponding author and its email address}
\corres{Isao T, Tokuda\\
\email{isao@fc.ritsumei.ac.jp}}

%%%% Abstract text to be placed here %%%%%%%%%%%%
\begin{abstract}
  A foremost challenge in modern network science is the inverse problem of reconstruction (inference) of coupling equations and network topology from the measurements of the network dynamics. Of particular interest are the methods that can operate on real (empirical) data without interfering with the system. One of such earlier attempts [Tokuda \textit{et al.}, PRL, 2007] was a method suited for general limit-cycle oscillators, yielding both oscillators' natural frequencies and coupling functions between them (phase equations) from empirically measured time series. 
The present paper reviews the above method in a way comprehensive to domain-scientists other than physics. It also presents applications of the method to 
(i) detection of the network connectivity, 
(ii) inference of the phase sensitivity function, 
(iii) approximation of the interaction among phase-coherent chaotic oscillators, 
(iv) experimental data from a forced Van der Pol electric circuit.
This reaffirms the range of applicability of the method for reconstructing coupling functions and makes it accessible to a much wider scientific community.
\end{abstract}
%%%%%%%%%%%%%%%%%%%%%%%%%%%

%%%%%%%%%% Insert the texts which can accomdate on firstpage in the tag "fmtext" %%%%%

\begin{fmtext}

\section{Introduction}
Complex networks are representations of complex systems, where nodes (vertices) represent system's units and links (edges) represent the interactions among those units \cite{newman,estrada,mason,alb}. The functioning of a real network is a cumulative effect of its structure (topology of connections among nodes/units) and dynamics (interactions/relationships among these nodes) \cite{mason,alb}. Hence, in models of real networks, nodes are often 
\end{fmtext}
%%%%%%%%%%%%%%% End of first page %%%%%%%%%%%%%%%%%%%%%
\maketitle
\noindent  
assumed to be (simple) systems with their inherent dynamics, whereas links mediate the dynamical coupling between the connected pairs of nodes. Using this paradigm, network science has made valuable contributions to all scientific disciplines that involve systems composed of many units, including biology, neuroscience, sociology, economics, \textit{etc.} \cite{newman,estrada,mason,alb,easley,boccaletti2014structure}.

To really grasp the functioning of a real network, we need information on both its structure and its dynamics. The inverse problem of reconstructing (or inferring) this information from the  empirical data is a foremost challenge in modern network science. Namely, understanding the inner connectivity patterns of real networks not only enables us to grasp their operations, but also helps in controlling and predicting their behavior \cite{blaha2011reconstruction,levnajic2011network,wang2011time,stankovski2012inference,kralemann2014reconstructing,levnajic2014untangling,timme2014revealing,han2015robust,nitzan2017revealing,wang2016data,leguia2017evolutionary,rosenblum2017reconstructing,simidjievski2018decoupling}. 

The problem of network reconstruction can be seen as composed of two parts. The first part is the reconstruction of network topology, where one tries to learn which pairs of nodes are connected and which are not. This can (in some cases) be done separately from the second part of the problem, which is the reconstruction of the coupling functions that dictate how the connected nodes interact. Of course, two parts of the problem are inherently related, but which one to tackle first depends on what data are available, what assumptions can be reasonably made about the system, and what exactly we wish to learn.

Numerous reconstruction methods have been proposed over the last decade, both in physics \cite{blaha2011reconstruction,levnajic2011network,wang2011time,stankovski2012inference,papana2013simulation,kralemann2014reconstructing,levnajic2014untangling,timme2014revealing,han2015robust,nitzan2017revealing,wang2016data,koutlis2016discrimination,leguia2017evolutionary,rosenblum2017reconstructing,simidjievski2018decoupling} and in computer science literature \cite{wang2016data,zanin2016combining,bridewell,dvzeroski2008equation,schmidt2009distilling,brunton2016discovering,simidjievski2016modeling,tanevski2016learning,tanevski2016process,vcerepnalkoski2012influence}. While some methods tackle only one of the two above mentioned parts of the problem \cite{stankovski2012inference}, other methods seek to address both parts at the same time. In physics literature, special emphasis is put on the methods aimed at oscillatory systems as the most researched paradigm of collective dynamics. This includes methods focused on either topology, coupling functions, or both \cite{blaha2011reconstruction,stankovski2012inference,levnajic2011network,rosenblum2017reconstructing}.

On a somewhat different front, research effort was devoted to the problem of estimation of phase variables and phase equations from the data \cite{tokuda2007inferring,kralemann2007uncovering,kralemann2008phase,cadieu2010phase,kralemann2011reconstructing,blaha2011reconstruction,zhu2013quantifying,kralemann2014reconstructing,pikovsky2018reconstruction,suzuki2018bayesian}. Namely, isolated units in many real systems exhibit oscillatory nature, in the sense that they can be well approximated as limit-cycle oscillators
(oscillator whose dynamics after transients reduces to periodic or quasi--periodic orbit). 
Researchers showed that, even if the oscillatory behavior is very stochastic, there are robust ways to extract a well defined phase variable for each network node, and hence reconstruct the phase equations that describe the system dynamics. This paradigm found applications in diverse domain sciences, notably biology and neuroscience, where many systems have this nature. Estimating phase equations, however, is nothing but reconstructing coupling functions from data. While such a reconstruction approach is valid only in the approximation of phase variables, these methods require very little additional assumptions about the system. This means they can be almost immediately applied to empirical data \cite{miyazaki2006determination,tokuda2007inferring,tokuda2010predicting,blaha2011reconstruction,zhu2013quantifying,stankovski2017coupling}.

For a system of phase equations, a standard way to construct 
the coupling function is to measure phase sensitivity 
function of an individual oscillator element and obtain 
the coupling function by averaging method that 
computes the amount of phase shift induced through 
interaction with another oscillator element
\cite{kiss2005predicting}.
However, a precisely measured phase sensitivity function is not 
always accessible, since it requires application of 
external perturbations to an individual oscillator, which cannot 
always be isolated from the rest of the system
\cite{galan2005efficient,ota2009map,nakae2010bayesian,hong2012efficient,goldberg2013spectral,saifee2015estimation,imai2017robust,cestnik2018inferring,okada2019acoustic,nakae2019statistical}.

On the other hand, as a non-invasive approach, the coupling 
function can be inferred directly from time-trace data 
measured from coupled oscillators
\cite{tokuda2007inferring,kralemann2007uncovering,kralemann2008phase,cadieu2010phase,tokuda2010predicting,kralemann2011reconstructing,blaha2011reconstruction,zhu2013quantifying,stankovski2017coupling,suzuki2018bayesian}.
One of them is a method by Tokuda \textit{et al.} \cite{tokuda2007inferring}: this approach utilized a multiple shooting method to realize robust parameter estimation of the coupling functions. The multiple shooting provides a general framework for fitting ordinary differential equations to recorded time-trace data. It is applicable to any system, where the dynamics of individual nodes can be approximated as those of limit-cycle oscillators, yielding both oscillators' natural frequencies and coupling functions between them (phase equations). Most importantly, the method was actually shown to operate very well with the data from a real experiment, which highlights its potential for practical use for physics problems and otherwise \cite{tokuda2007inferring,tokuda2010predicting,tokuda2013detecting}.

The contribution of the present paper is two-fold. First, we review this method in a way that is understandable and approachable to communities outside physics. With this, we hope to make our method more useful to field such as biology and neuroscience, for which it was originally intended. 
Second, we show and discuss applications of this method, specifically: 
(i) we utilize the estimated coupling function for detecting the connectivity of oscillator networks,
(ii) the method is extended to inference of the phase sensitivity function, which is vital for phase equations, 
(iii) the coupling function is estimated for coupled chaotic oscillators to demonstrate how well the phase model approximates chaotic phase synchronization, 
(iv) using an experimental data from a system of Van der Pol electric circuits, we show how the method can be applied to real data.

The rest of the paper is organized as follows. In the next section, we review the original method in a comprehensive way. 
In section 3, we discuss the problem of inferring the network connectivities.
In section 4, we present further applications mentioned above.
In the last section, we discuss our findings and lay out perspectives for future work.

\section{The Original Method}
In this section, we describe the original method in a more 
comprehensive way than the original literature
\cite{tokuda2007inferring}
and show how it works for the case of 
coupled FitzHugh-Nagumo oscillators. 

\subsection{Multiple Shooting Method}
Our approach is based on the multiple shooting method, which 
has been developed in physics and engineering to provide 
a general framework for fitting ordinary differential 
equations to recorded time series
\cite{baake1992fitting}. 
The methodology is applicable to a situation, where the system 
equations are known \textit{a priori}.
When the equations and the recorded data are in a good
quantitative agreement, unknown parameters of the system
can be precisely estimated as follows.

We consider a nonlinear system 
\begin{eqnarray}
{\dot{\bm{x}}} &=& \bm{F} (\bm{p}, \bm{x}), 
\label{eq:original_ode_msm}
\end{eqnarray}
where $\bm{x}$, $\bm{p}$  and $\bm{F}$ represent state variables,
parameters, and autonomous dynamics of the system, respectively.
The system may generate nonlinear dynamics such as limit cycles, 
torus, strange attractors, and transient dynamics to these attractors.
The equation~(\ref{eq:original_ode_msm}) may describe a variety of 
systems of interest in science and engineering such as
electric circuits and lasers. 
Empirical data consists of oscillators' states measured as
${\{} \bm{x} ( n {\Delta}t)): n=1,{\cdots},M {\}}$
(${\Delta}t$: sampling time, $M$: data points).
This corresponds to an experimental situation, in which
the system state (\textit{e.g.}, current and voltage of
electric circuits, laser, \textit{etc.}) is fully recorded.
Then the parameter values $\bm{p}$ that underly the measurement 
data can be estimated by fitting the original 
equations~(\ref{eq:original_ode_msm}) to the recording data. 
First, time evolution of the original 
equations~(\ref{eq:original_ode_msm})
starting from an initial condition
$\bm{x}(0)$ is denoted by
$\bm{\phi}^{t} (\bm{x}(0), \bm{p})$.
Then, at each sampling time $t = i {\Delta}t$,
the equations must satisfy the boundary conditions:
$\bm{x}( (n+1) {\Delta}t )$ $=$
$\bm{\phi}^{{\Delta}t} (\bm{x}( n {\Delta}t ), \bm{p})$.
With respect to the unknown parameters $\bm{p}$,
the nonlinear equations are solved by the 
generalized Newton method \cite{press2007numerical}.
To compute the gradients
${\partial}{\phi}_{i}/{\partial}p$,
which are needed for the Newton method,
variational equations 
$\frac{d}{dt} (\frac{{\partial}{\phi}_{i}}{{\partial}p_{j}}) =
\frac{{\partial}f_{i}}{{\partial}p_{j}} +
\sum_{k=1}^{N}
\frac{{\partial}f_{i}}{{\partial}{\phi}_{k}}
\frac{{\partial}{\phi}_{k}}{{\partial}p_{j}}$
are solved simultaneously, 
where $f_{i}$ represents $i$th equation of the
original dynamics~(\ref{eq:original_ode_msm}) as 
${\dot{x}}_{i} = f_{i} (\bm{x}, \bm{p})$.
The evolution function ${\phi}^{t}$ as well as the
variational equations are integrated numerically, using 
whichever integration scheme (\textit{e.g.}, 
4th--order \textit{Runge-Kutta}).
It has been shown that, when the equations and the 
experimental data are in good quantitative agreement, 
the unknown parameters can be precisely estimated for 
real--world systems including electric circuits and lasers.
All steps in the above computational procedure can be 
realized relatively easily with standard programming 
knowledge.

\subsection{Problem and Method}
Equipped with the knowledge of multiple shooting method, 
we now explain in detail how it can be utilized for inferring 
the coupling functions. We begin by considering a network composed 
of interacting oscillator elements.
In biology, such systems include a network of circadian cells
in the suprachiasmatic nucleus \cite{yamaguchi2003},
brain network composed of many spiking neurons
\cite{galan2005efficient,hong2012efficient},
cardiac muscle cells in the heart
\cite{verheijck1998pacemaker},
\textit{etc.}
In terms of nonlinear dynamics, such systems are described
as a system of $N$ coupled limit cycle oscillators:
\begin{eqnarray}
{\dot{\bm{x}}}_{i} &=& \bm{F}_i (\bm{x}_{i}) + \frac{C}{N} 
\sum_{j=1,j{\neq}i}^{N} T_{i,j} 
\bm{G} (\bm{x}_{i}, \bm{x}_{j}). 
\label{eq:original_ode}
\end{eqnarray}
Here, $\bm{x}_i$ and $\bm{F}_i$ ($i=1,2,{\cdots},N$) 
represent state variables and autonomous dynamics of the 
$i$th oscillator element, respectively.
While $\bm{G}$ represents an interaction function between 
$i$th and $j$th oscillators, strength of their interaction 
is determined by the coupling constant $C$. 
The matrix ${\{} T_{i,j} {\}}$ describes connectivity 
between the oscillator elements.
For simplicity, we suppose that the connection matrix 
is composed of zero-or-unity elements 
(\textit{i.e.}, $T_{i,j}=0$ or $1$).
We assume that, without coupling (\textit{i.e.}, $C=0$), 
individual systems
(\textit{i.e.}, ${\dot{\bm{x}}}_{i} = \bm{F}_i (\bm{x}_{i})$)
generate periodic oscillations, after transients.
Such closed trajectories in phase space are called 
\textit{limit cycles}, 
which have intrinsic periods of ${\tau}_{i}$.
The equation~(\ref{eq:original_ode}) describes, to a good 
approximation, a variety of systems of interest in biology 
and neuroscience. Then the theory of phase reduction 
\cite{winfree2001,kuramoto1984}
states that, as far as the coupling strength $C$ is weak 
enough, the network dynamics can be reduced to the following 
phase equations:
\begin{eqnarray}
{\dot{\theta}}_{i} &=& 
{\omega}_{i} + \frac{C}{N} 
\sum_{j=1,j{\neq}i}^{N} T_{i,j} 
\bm{Z} ({\theta}_{i}) 
\bm{G} ({\theta}_{i}, {\theta}_{j})
= {\omega}_{i} + \frac{C}{N} 
\sum_{j=1,j{\neq}i}^{N} T_{i,j} H ({\theta}_{j} - {\theta}_{i}),
\label{eq:phase_dynamics}
\end{eqnarray}
where ${\theta}_i$ represents phase of the $i$th oscillator and
${\omega}_{i}$ gives natural frequency of the $i$th 
oscillator (\textit{i.e.}, ${\omega}_{i}=2{\pi}/{\tau}_{i}$).
$\bm{Z}$ stands for phase sensitivity function
(also called ``infinitesimal phase response curve''),
which determines the amount of phase shift induced by 
the interaction $\bm{G}$ with other oscillators
(we will here not go in the detail of how 
equation~(\ref{eq:phase_dynamics}) is obtained; 
an interested reader can refer to 
\cite{winfree2001,kuramoto1984,pikovsky2003synchronization}).
By averaging approximation \cite{kuramoto1984}, which 
integrates one cycle of the phase sensitivity function 
$\bm{Z}$ with the interaction function $\bm{G}$,
the coupling function is derived as
$H({\theta}_{i}-{\theta}_{k}) = 
\frac{1}{2{\pi}} 
{\int}_{0}^{2{\pi}} \bm{Z} ({\theta}_{i}+{\theta}') 
\bm{G} ({\theta}_{i}+{\theta}', {\theta}_{k}+{\theta}')
d{\theta}'$.
Transformation of the original 
equations~(\ref{eq:original_ode}) to 
(\ref{eq:phase_dynamics}) provides a significant reduction 
in system's dimensionality, in the sense that 
the original state variables $\bm{x}_i$, which can be
high-dimensional, are represented only by the single 
phase variable ${\theta}_i$.
This substantially simplifies the system's modeling and 
enables its identification in a straightforward fashion. 

The individual oscillator states are simultaneously measured as
${\{} {\xi}_i (n {\Delta}t) = {\eta} (\bm{x}_i ( n {\Delta}t)): 
n=1,{\cdots},M {\}}_{i=1}^{N}$
(${\eta}$: observation function, ${\Delta}t$: sampling time,
$M$: data points).
This corresponds to an experimental situation, under which
states of individual oscillators (\textit{e.g.}, gene expression 
levels of individual cells, membrane potentials of neurons,
\textit{etc.}) are recorded simultaneously. 
Our purpose is to infer the phase equations from these 
measurement data under the conditions that the underlying 
system equations~(\ref{eq:original_ode}) are unknown.

The phase dynamics can be reconstructed via the following steps.
\begin{enumerate}
\item From the measured data ${\{}x_{i}(t){\}}$, 
phases are extracted as 
${\theta}_{i} (t) = 2{\pi}k + 2{\pi}(t-t_{k})/(t_{k+1}-t_{k})$,
where $t_{k}$ represents the time, at which $i$th signal
takes its $k$th peak and $t_{k}{\leq}t<t_{k+1}$
\cite{pikovsky2003synchronization}.
Note that this method is limited to the case of simple 
waveform, where a single peak appears during one oscillation 
cycle.

\item Fit the phase equations:
\begin{eqnarray}
{\dot{\theta}}_{i} &=& \tilde{{\omega}}_{i} \; + \;
\frac{C}{N} \sum_{j=1}^{N} \tilde{T}_{i,j} \tilde{H}
({\theta}_{j} - {\theta}_{i}) 
\label{eq:model_phaseeq1}
\end{eqnarray}
to the phase data ${\{} {\theta}_{i} (t) {\}}$.
Here, ${\{} \tilde{{\omega}}_{i} {\}}$ represent 
approximate values for the natural frequencies.
The coupling function $\tilde{H}$, which is 
in general nonlinear and periodic with respect to $2\pi$, 
is approximated by a \textit{Fourier} series of pre-selected 
order $D$ as
$\tilde{H} ({\Delta}{\theta})$ $=$
$\sum_{j=1}^{D} a_{j} {\sin} j {\Delta}{\theta}
+ b_{j} ( {\cos} j {\Delta}{\theta} - 1)$.
For simplicity, we consider a specific type of coupling,
under which the interaction disappears as the 
phase difference becomes zero, {\it i.e.,} 
$\tilde{H}(0)=0$. 
This type of coupling arises quite often in diffusively 
coupled oscillator networks 
\cite{kuramoto1984,acebron2005kuramoto}.
[Although general coupling can be also considered, more
than one data sets associated with different coupling
strength are required to avoid parameter redundancy. 
As a simplified demonstration, this study deals with this
specific coupling.]

The unknown parameters 
$\bm{p} = {\{} \tilde{\omega}_{i}, a_j, b_j {\}}$
are now estimated by the above described multiple-shooting method
(the connection matrix $\tilde{T}_{i,j}$ and the coupling 
strength $C=0$ are supposed to be known here).

\item
To avoid over-fitting of the coupling function, 
cross-validation is utilized 
to determine the optimal number of \textit{Fourier} 
components $D$ \cite{stone1977asymptotic}.
We divide the measurement data into two parts. 
For the first half data, the parameter values $\bm{p}$ 
are estimated. Then, the estimated parameters are 
applied to the latter half data and measure the error
$e_{cv} = \sum_n {\vert} \bm{\theta}( (n+1) {\Delta}t ) - 
\bm{\phi}^{{\Delta}t} (\bm{\theta}( n {\Delta}t ), \bm{p})
{\vert}^{2}$, where 
$\bm{\phi}^{{\Delta}t} (\bm{\theta}( n {\Delta}t ), \bm{p})$
represents ${\Delta}t$-time further state of the phase 
dynamics~(\ref{eq:model_phaseeq1}) starting from an 
initial condition $\bm{\theta}( n {\Delta}t )$.
The order number $D$ providing the minimum error is 
considered as the optimum.
\end{enumerate}

\subsection{Application to Coupled FitzHugh-Nagumo Oscillators}
To illustrate how the method described above works, 
we apply the multiple shooting to a prototypical 
example of weakly coupled limit cycle oscillators.
In the original study \cite{tokuda2007inferring}, 
coupled R\"{o}ssler oscillators were analyzed.
As another yet challenging example, which has more 
complex shape of coupling function due to the nature of 
relaxation oscillations, we consider the following 
network of FitzHugh-Nagumo (FHN) oscillators
\cite{fitzhugh1961impulses,nagumo1962active}:
\begin{eqnarray}
{\dot{v}}_{i} &=& {\alpha}_{i} ( v_{i} - v_{i}^{3}/3 - w_{i} + I )
+ \frac{C}{N} \sum_{j=1}^{N} T_{i,j} ( x_{j} - x_{i} ), 
\label{eq:coupled_FHN1}\\
{\dot{w}}_{i} &=& {\alpha}_{i} {\epsilon} ( v_{i} + a - b y_{i} ),
\label{eq:coupled_FHN2}
\end{eqnarray}
where $i=1,{\cdots},N$.
The system of FitzHugh-Nagumo oscillators can be seen 
as a simple model for interacting neurons. 
Under the parameter setting of 
$a = 0.7$, $b = 0.8$, ${\epsilon} = 0.08$, $I = 0.8$,
individual FHN oscillators (without coupling $C=0$)
gives rise to limit cycles of slow-fast type. 
Inhomogeneity parameters, which control natural periods 
of the individual oscillators, were set as
${\alpha}_{i}=1+(i-1){\Delta}{\alpha}$ ($i=1,{\cdots},N$),
where ${\alpha}_{i}=1$ yields natural oscillation period 
of $36.5$.

We started with the case of $N=16$. 
We consider all-to-all coupling matrix ($T_{i,j}=1$). 
The level of inhomogeneity was set to ${\Delta}{\alpha}=0.01$.
The multivariate data ${\{} x_i (t) {\}}_{i=1}^{16}$ were 
recorded at a coupling strength of $C=0.01$, 
which is in a non-synchronized regime.
The sampling interval was set to be ${\Delta}t = 0.004$.
Then, the phases $\{ {\theta}_{i} (t) \}$ were
extracted and down-sampled to a sampling interval of
${\Delta}t = 1000 {\cdot} 0.004$.
Total of $500$ data points have been collected for 
the parameter estimation.
As an initial condition, the unknown parameter 
values were all set to be zero, {\it i.e.}, 
${\tilde{\omega}}_{i}=0$, $a_j = b_j = 0$.
The $500$ data were divided into $250$ and $250$,
which were used for the parameter estimation and
the cross-validation error $e_{cv}$, respectively.
By varying the number of \textit{Fourier} components 
from $D=1$ to $D=10$,
the optimal value was found to be $D=7$ by the
cross-validation test.

\begin{figure}[tb]
\begin{center}
{\scriptsize (a)}\hspace{2.5mm}
  \includegraphics[width=60mm]{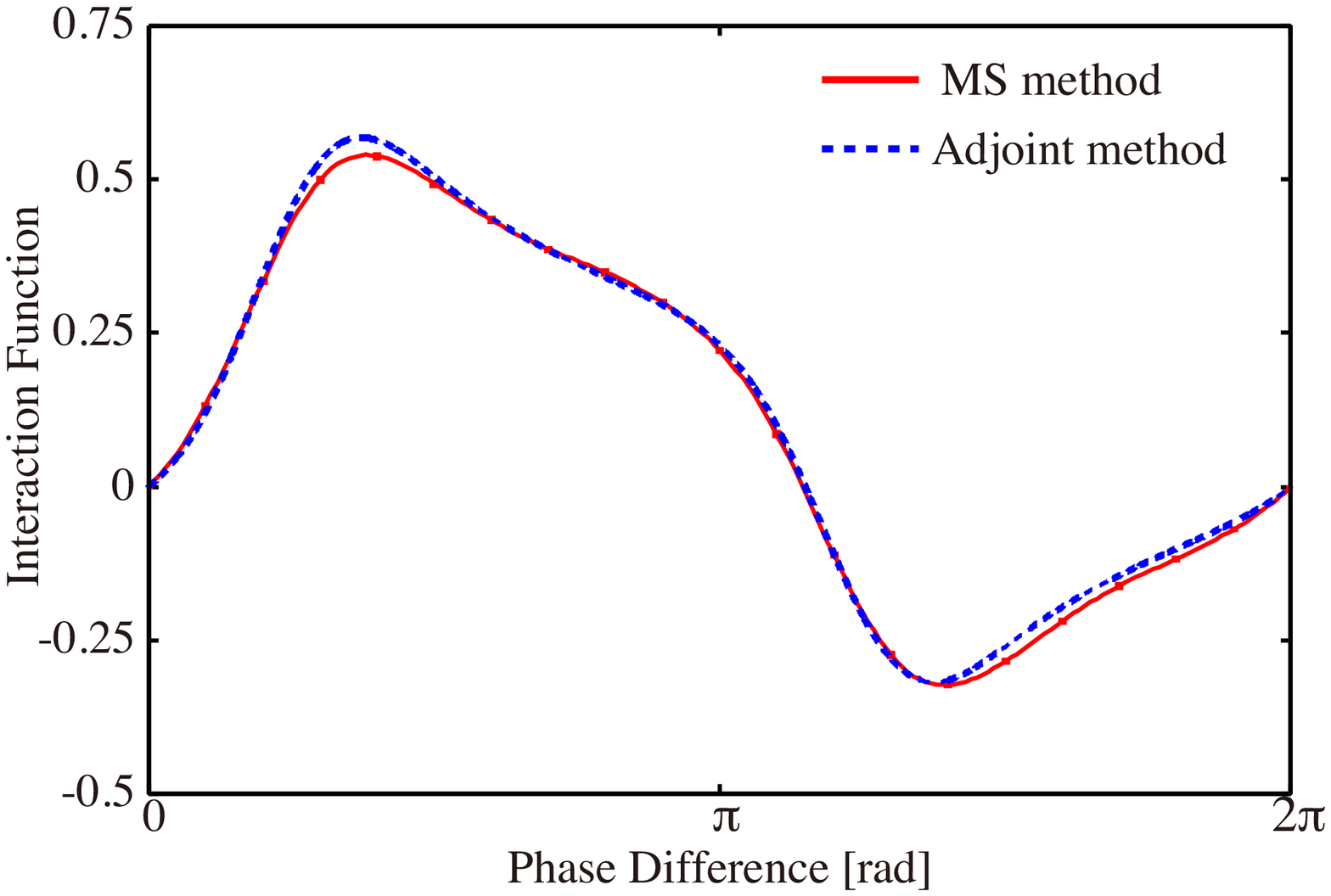}
{\scriptsize (b)}\hspace{2.5mm}
  \includegraphics[width=60mm]{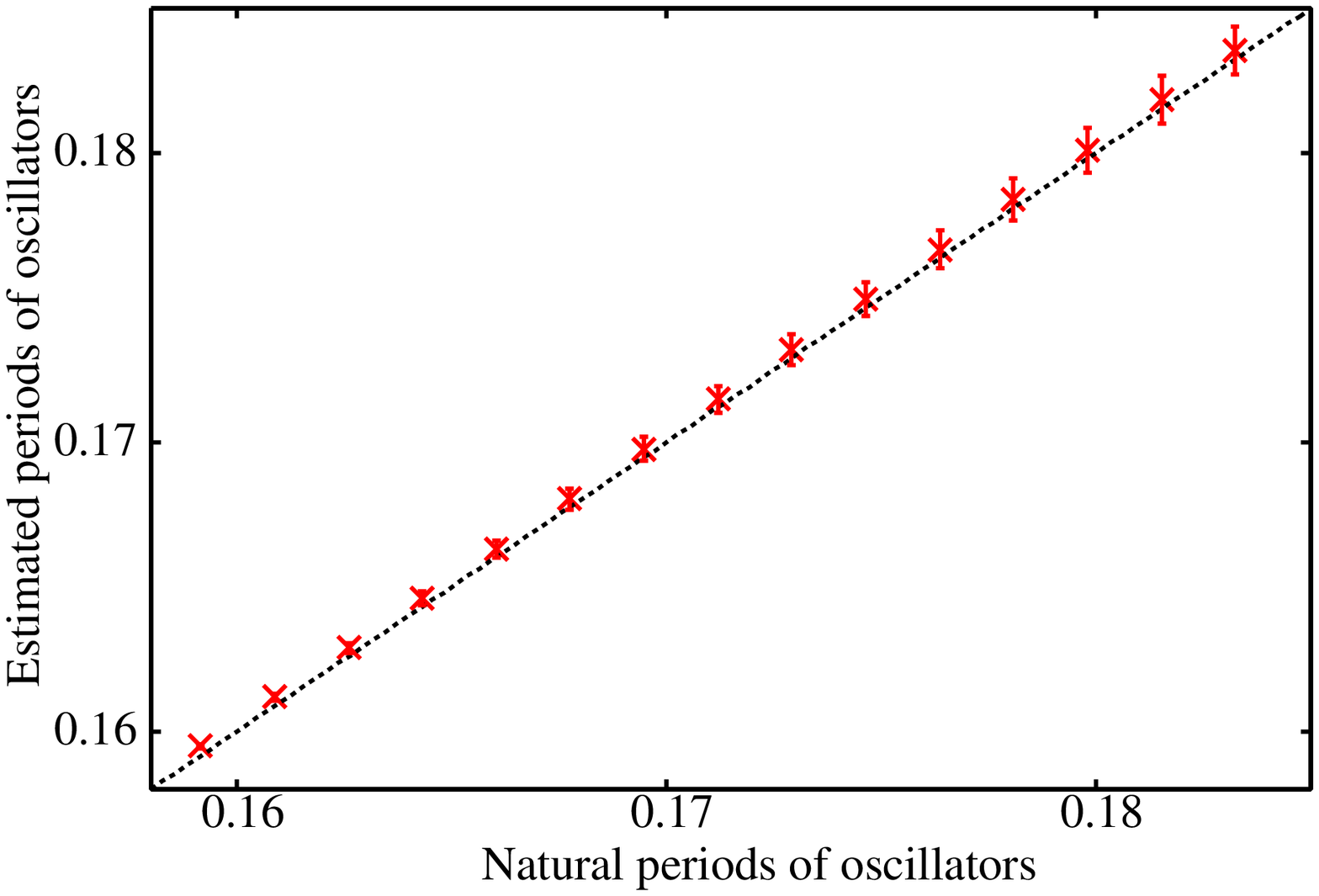}
\vspace{5mm}\\
{\scriptsize (c)}\hspace{2.5mm}
  \includegraphics[width=60mm]{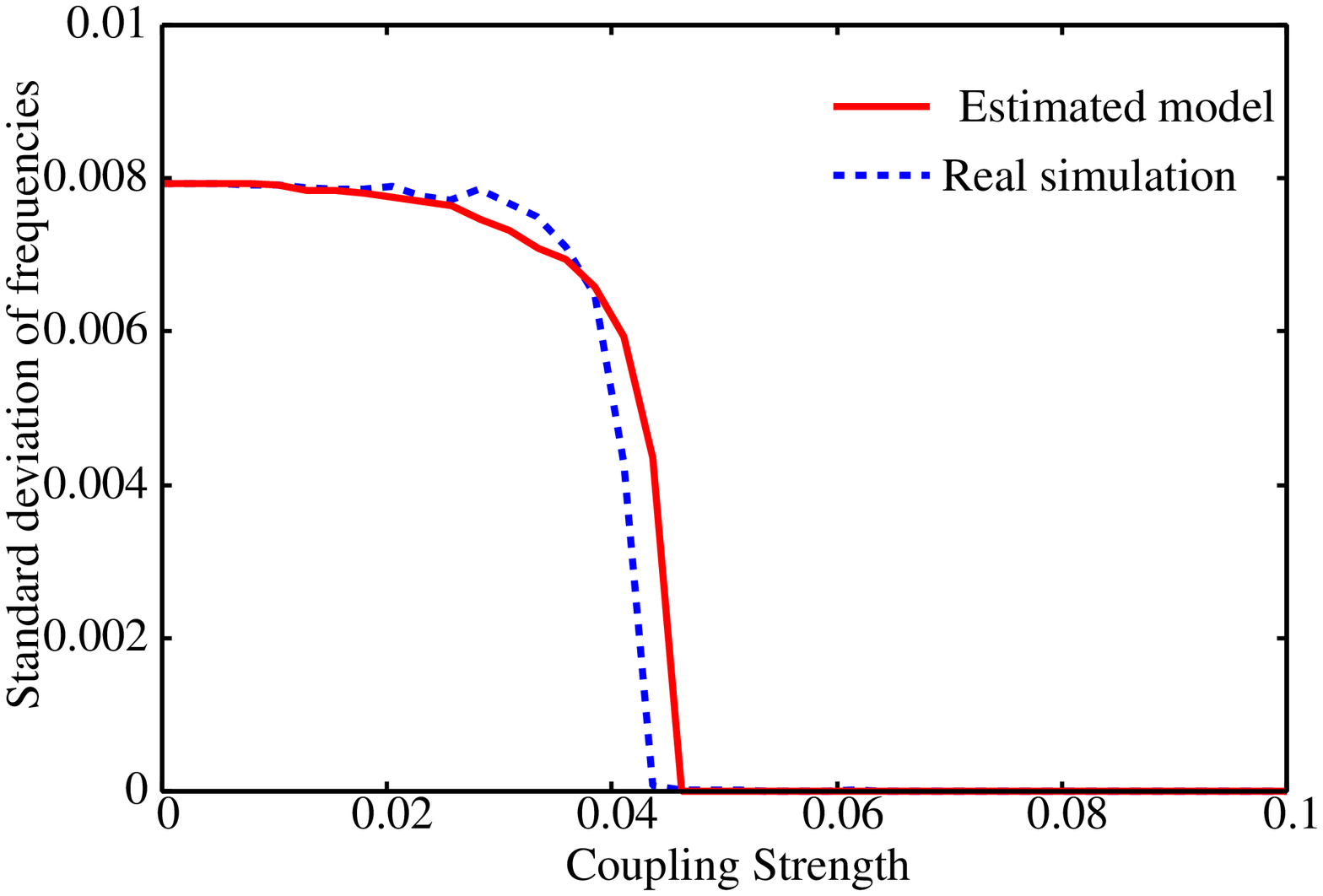}
{\scriptsize (d)}\hspace{2.5mm}
  \includegraphics[width=60mm]{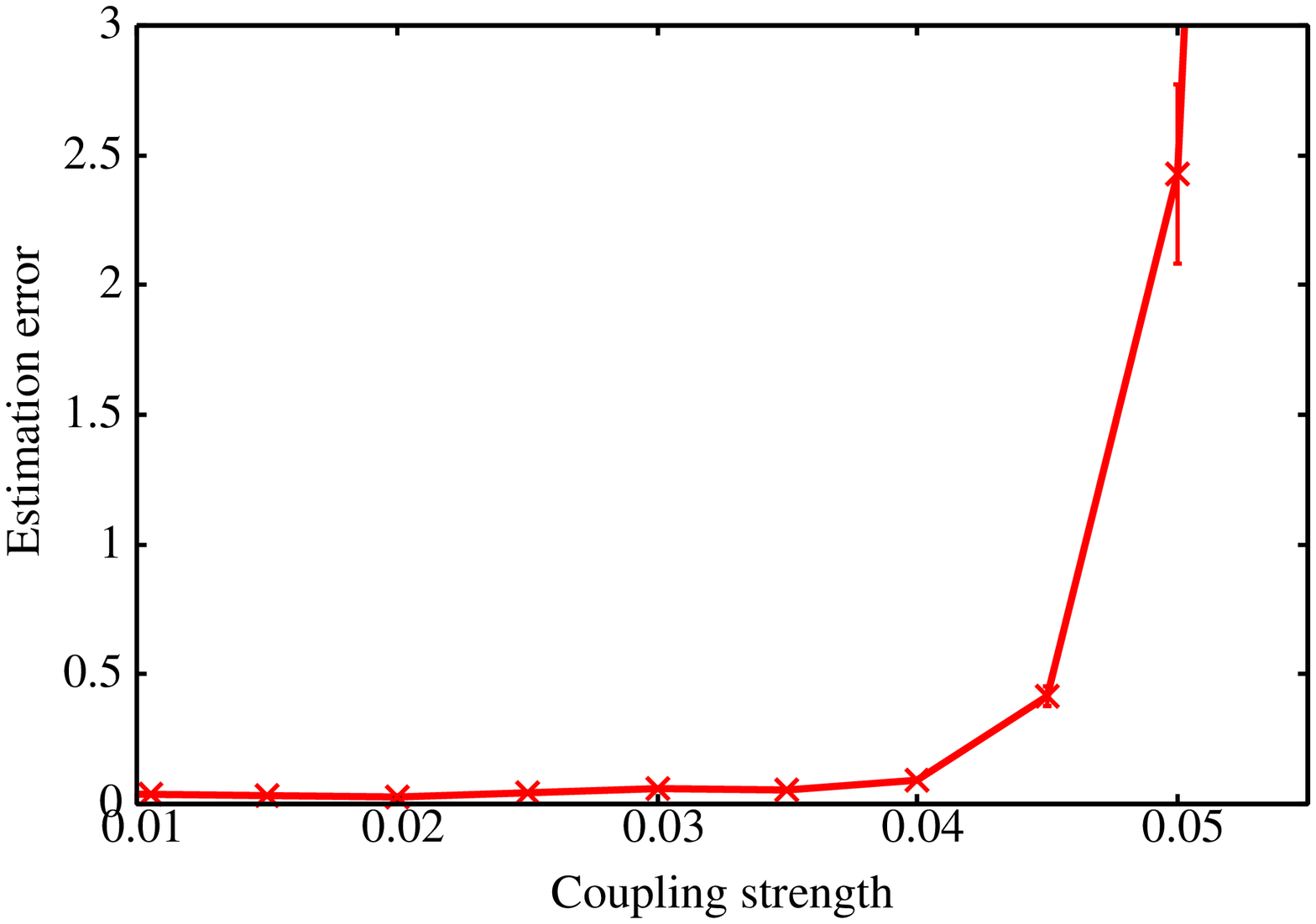}
\vspace{5mm}\\
{\scriptsize (e)}\hspace{2.5mm}
  \includegraphics[width=60mm]{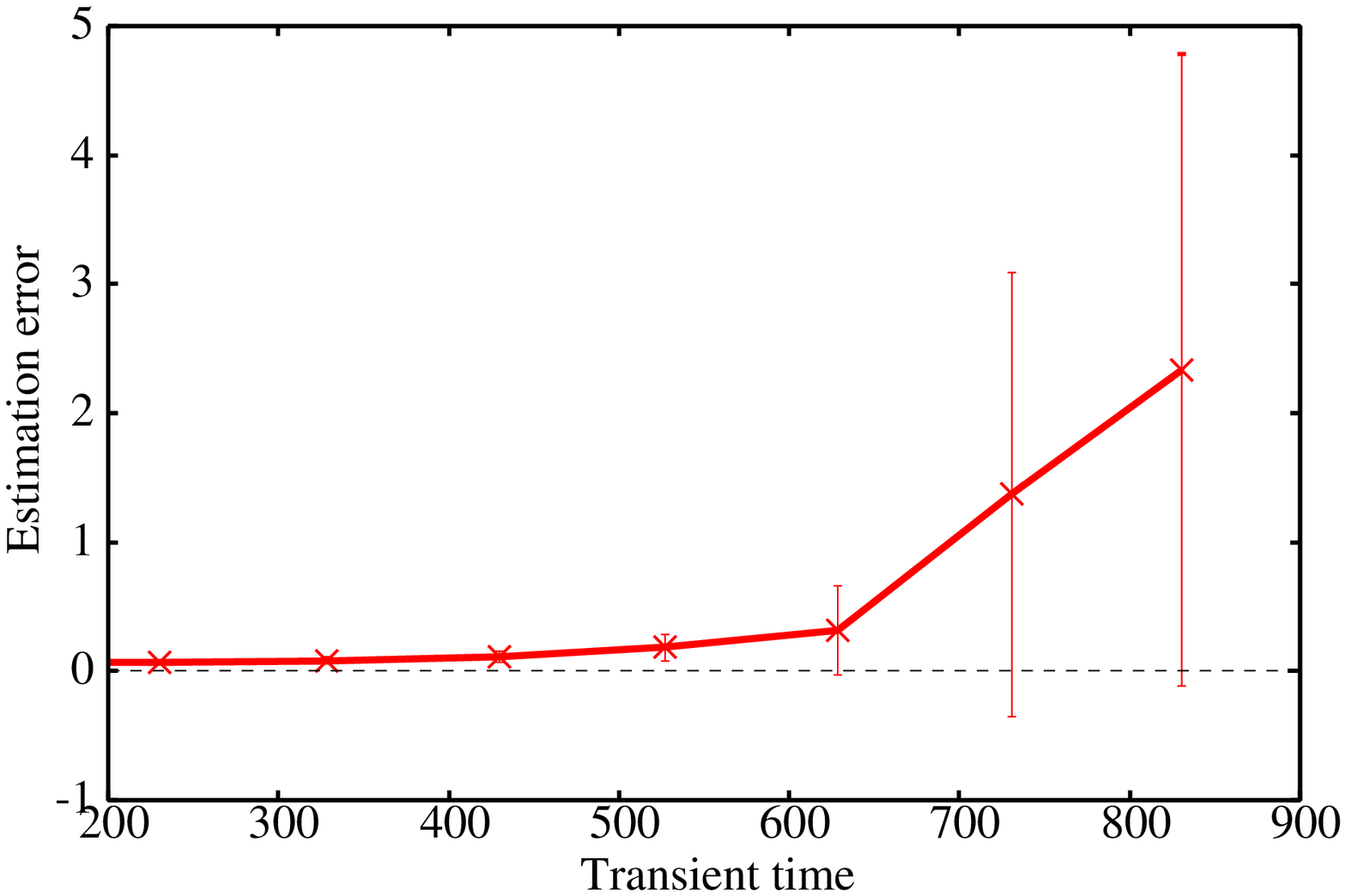}
{\scriptsize (f)}\hspace{2.5mm}
  \includegraphics[width=60mm]{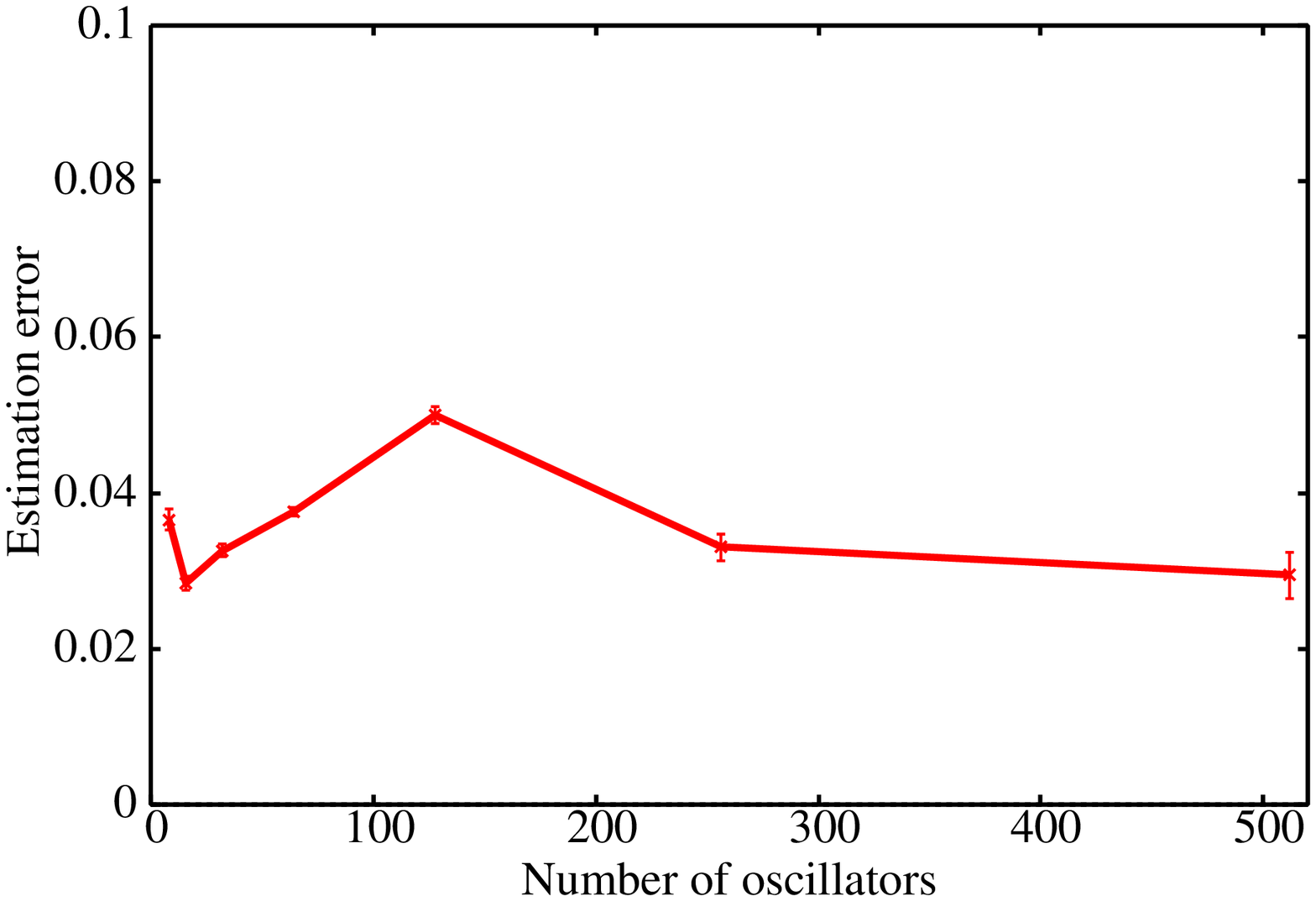}
  \caption{Results for a network of $N=16$ FHN oscillators.
(a) Coupling functions $\tilde{H} ({\Delta}{\theta})$
estimated by the present method (solid red line) 
and the adjoint method (dashed blue line).
(b) Estimated natural frequencies (ordinate)
${\{} {\omega}_{i} {\}}_{i=1}^{16}$ of FHN oscillators 
\textit{vs.}
those obtained from non-coupled simulation (abscissa).
(c) Synchronization diagrams of the estimated model 
(solid red line) and the original coupled oscillators
(dashed blue line).
(d) Dependence of estimation error on the coupling
strength $C$ used to generate multivariate data.
The estimation error $e$ is defined as the deviation
of the estimated coupling function from 
the one computed by the adjoint method.
(e) Dependence of the estimation error on the transient
time, after which the multivariate data were sampled.
The coupling strength was set to $C=0.05$.
(f) Dependence of the estimation error on the number of
oscillators $N$. 
}
\label{fig:16Oscillators}
\end{center}
\end{figure}

The coupling function $\tilde{H} ({\Delta}{\theta})$
estimated by the present method is in a good agreement 
with the one computed by the adjoint method 
\cite{ermentrout1996type}
(Fig.~\ref{fig:16Oscillators}a).
The error-bars were computed from inverse of the 
Hessian matrix of the squared error function,
under the assumption that the phase data contain 
uncorrelated observational noise 
\cite{horbelt2001identifying}.
The estimated natural frequencies are distributed on 
a diagonal line with the true frequencies obtained
from simulations of the individual (isolated) FHN 
oscillators (Fig.~\ref{fig:16Oscillators}b).
Using the estimated phase equations, synchronization 
diagram of the original coupled FHN oscillators can 
be recovered, where the onset of synchronization was
predicted at $C=0.046$, which is very close to the
real onset of $C=0.044$
(Fig.~\ref{fig:16Oscillators}c). 

Next, we show how the estimation depends upon the 
problem setting. 
The primary factor that influences the estimation 
results is the coupling strength $C$ used to generate 
the time series.
Fig.~\ref{fig:16Oscillators}~(d) shows dependence of 
estimation error on the coupling strength.
The estimation error $e_{cf}$ is evaluated as 
deviation of the estimated coupling function 
$\tilde{H}_{s} ({\Delta}{\theta})$
from the one
$\tilde{H}_{p} ({\Delta}{\theta})$
estimated by the adjoint method,
\textit{i.e.}, 
\begin{eqnarray}
e_{cf} = 
\frac{\sqrt{ \int_{0}^{2\pi} {\{}
\tilde{H}_{s} ({\Delta}{\theta}) - 
\tilde{H}_{p} ({\Delta}{\theta}) {\}}^{2}
d {\Delta}{\theta} }}{
\sqrt{
\int_{0}^{2\pi} {\{}
\tilde{H}_{p} ({\Delta}{\theta}) - 
{\langle} \tilde{H}_{p} {\rangle} {\}}^{2} 
d {\Delta}{\theta} }},
\label{eq:estimationerror}
\end{eqnarray}
where the denominator represents normalization factor 
and
${\langle} \tilde{H}_{p} {\rangle} =
(1/({2\pi}))\int_{0}^{2\pi} 
\tilde{H}_{p} ({\Delta}{\theta}) d{\Delta}{\theta}$.
As the coupling strength is located close to the onset 
of synchronization, the estimation error increases 
significantly.
Under the synchronized state, phase differences between 
the oscillators do not change in time 
${\Delta}{\theta}=const$, providing no information about 
the phase interaction. Increase in estimation error due
to synchronized data is therefore reasonable. 

Even in a synchronized regime, the coupling function 
can be recovered from transient data, during which
phase differences evolve
(transient data often reveals far more information about 
the underlying system, since it is recorded before the 
system 'settled' into its dynamical equilibrium).
To show this, the multivariate data were recorded after
discarding only a short duration of transient process
that starts from a random initial condition. 
Transient data (time interval of $40$) were collected before 
the system reached the final synchronized state.
$20$ sets of such data were used for the parameter estimation.
Fig.~\ref{fig:16Oscillators}~(e) shows dependence of the 
estimation error on the transient duration.
Note that the coupling strength is set to $C=0.05$, that
is in a synchronized regime.
Although the error increases as the transient duration is 
increased, relatively good estimation was realized for 
a short transient time. This suggests that, even the system 
is in synchrony with a moderate coupling, application of 
perturbation that kicks the system out of synchrony is 
an efficient way of inferring the underlying phase dynamics.

Fig.~\ref{fig:16Oscillators}~(f) shows dependence of the 
estimation error on the network size $N$, varied from 
$8$ to $512$. 
The level of inhomogeneity was set to ${\Delta}{\alpha}=0.16/N$.
The multivariate data ${\{} y_i (t) {\}}_{i=1}^{N}$ were 
recorded at a coupling strength of $C=0.02$, 
which corresponds to non-synchronized regime.
Other settings were the same as those in the case of $N=16$. 
Surprisingly, the estimation error remains in a low level.
Even for $N=512$, the coupling function 
$\tilde{H} ({\Delta}{\theta})$
has been precisely estimated, while 
the estimated natural frequencies 
${\{} {\omega}_{i} {\}}_{i=1}^{512}$ 
are consistent with those obtained from the non-coupled 
simulations. 
This suggests that the system size does not pose a major 
limit on the estimation of phase dynamics as far as 
the data contain non-synchronized phase information. 

Although the coupling function has been reliably estimated
for networks with all-to-all connections ($T_{i,j}=1$), 
the estimation error may increase when oscillators are
heterogeneously connected to each other.
We deal with such situations in the next section.

\section{Application to Network Inference}
Although we have dealt with the case that all oscillator 
elements are connected to all the others in the previous
section, heterogeneous connections are more common in 
nature and engineering.
As another challenge of our technique \cite{tokuda2013detecting},
this section discusses a problem of inferring connectivity 
of the oscillator network from the measured time series.
Numerous approaches have been proposed up to date using
information transfer \cite{schreiber2000measuring}, 
mutual predictability \cite{rosenblum2001detecting}, 
recurrence properties \cite{romano2007estimation},
permutation-based asymmetric association measure 
\cite{hempel2011inner},
index for partial phase synchronization 
\cite{schelter2006partial,nawrath2010distinguishing,wickramasinghe2011phase},
and graph theory \cite{runge2012escaping}.
Response properties of the network dynamics to external 
stimuli have been also exploited 
\cite{timme2007revealing,levnajic2011network}.
For weakly coupled limit cycle oscillators, to which 
phase reduction is applicable, the phase modeling approach
is again quite effective for detecting the network topology
\cite{tokuda2013detecting,kralemann2014reconstructing,stankovski2015coupling,stankovski2016alterations,stankovski2017neural}.

In our approach \cite{tokuda2013detecting}, 
the multiple-shooting method is again applied to fit 
the phase equations~(\ref{eq:model_phaseeq1})
to the phase data ${\{} {\theta}_{i} (t) {\}}$. 
The fitting procedure is the same as before except that
the connection matrix is estimated as the unknown parameters 
$\bm{p} = {\{} \tilde{T}_{i,j} {\}}$.
For simplicity, the coupling function $\tilde{H}$ and
the natural frequencies ${\{} {\omega}_{i} {\}}_{i=1}^{N}$ 
of the oscillator elements were assumed to be known
(the general case that both coupling function and connection 
matrix are unknown has been dealt with in the previous 
study \cite{tokuda2013detecting}).
As coefficients ${\{} a_j, b_j {\}}$ for the coupling function,  
the ones estimated in the previous section were utilized. 
Natural frequencies ${\{} {\omega}_{i} {\}}_{i=1}^{N}$ were
also obtained from the simulations of non-coupled 
original equations.

As the target system, the network of FHN 
oscillators~(\ref{eq:coupled_FHN1}),(\ref{eq:coupled_FHN2})
were studied.
For a network of four ($N=4$) oscillators, two defects
were introduced to the connection matrix
as $T_{3,1}=T_{4,1}=0$
(here, \textit{defect} means that one oscillator is not 
connected to another). 
The level of inhomogeneity was set to ${\Delta}{\alpha}=0.04$,
whereas the coupling strength was $C=0.02$, \textit{i.e.},
in a non-synchronized regime.
As in the previous section, total of $500$ data points 
(sampling time: $4$) have been collected.
By the multiple-shooting method, the connection matrix 
was estimated as follows.
\begin{eqnarray}
\left(
\begin{array}{llll}
                & \tilde{T}_{1,2} & \tilde{T}_{1,3} & \tilde{T}_{1,4} \\
\tilde{T}_{2,1} &                 & \tilde{T}_{2,3} & \tilde{T}_{2,4} \\
\tilde{T}_{3,1} & \tilde{T}_{3,2} &                 & \tilde{T}_{3,4} \\
\tilde{T}_{4,1} & \tilde{T}_{4,2} & \tilde{T}_{4,3} \\
\end{array}
\right) &=& 
\left(
\begin{array}{llll}
               & 1.11{\pm}0.01 & 1.08{\pm}0.01 & 1.06{\pm}0.01\\
 1.06{\pm}0.02 &               & 1.04{\pm}0.02 & 0.97{\pm}0.01\\
-0.02{\pm}0.01 & 1.05{\pm}0.01 &               & 1.03{\pm}0.01\\
 0.04{\pm}0.01 & 0.98{\pm}0.01 & 1.03{\pm}0.01 
\end{array}
\right) .
\nonumber
\end{eqnarray}
We see that the two defects 
($\tilde{T}_{3,1}$, $\tilde{T}_{4,1}$)
were precisely identified
as small values, whereas other matrix elements pointed 
almost unity.

\begin{figure}[tb]
\begin{center}
{\scriptsize (a)}\hspace{2.5mm}
  \includegraphics[width=60mm]{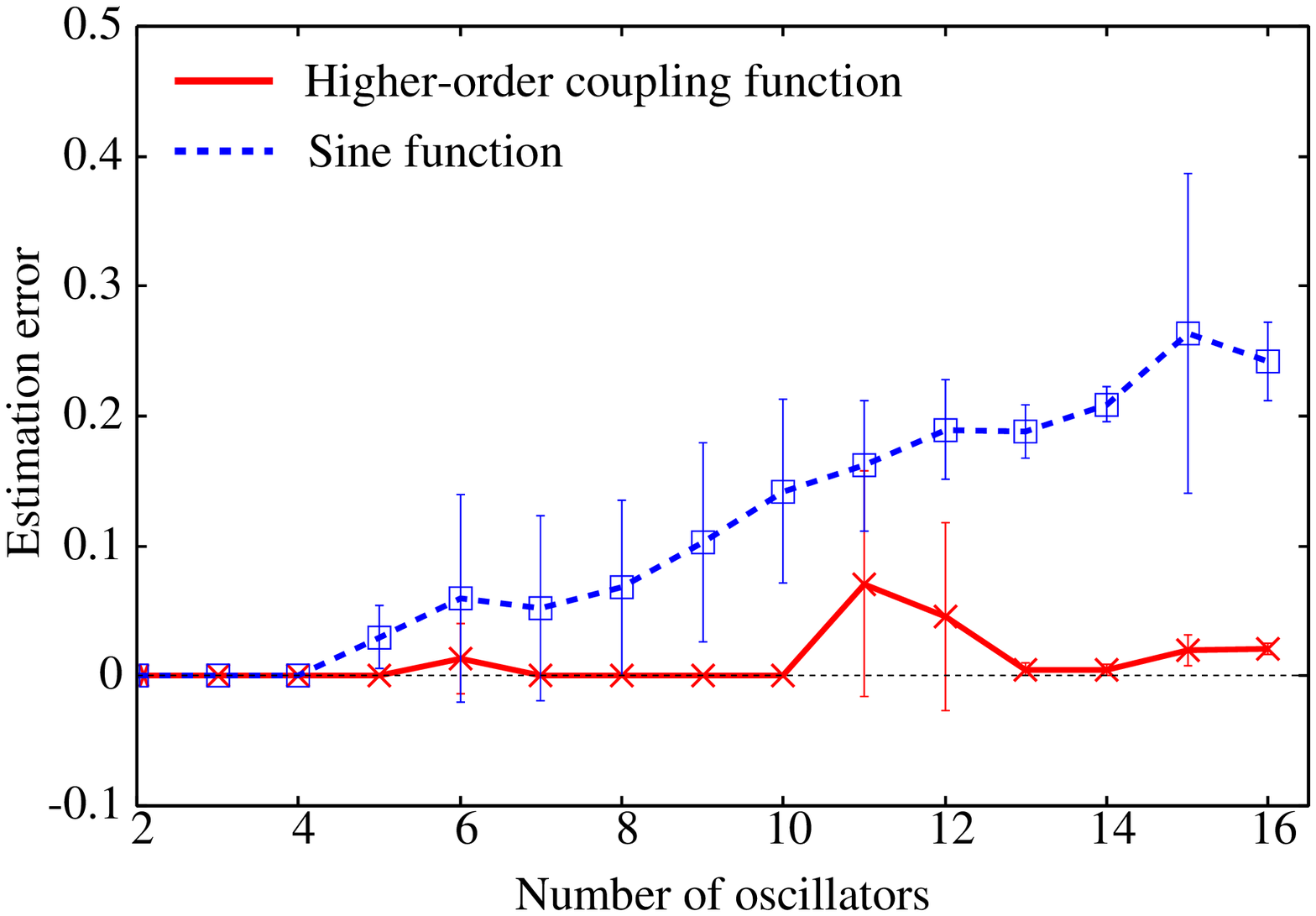}
{\scriptsize (b)}\hspace{2.5mm}
  \includegraphics[width=60mm]{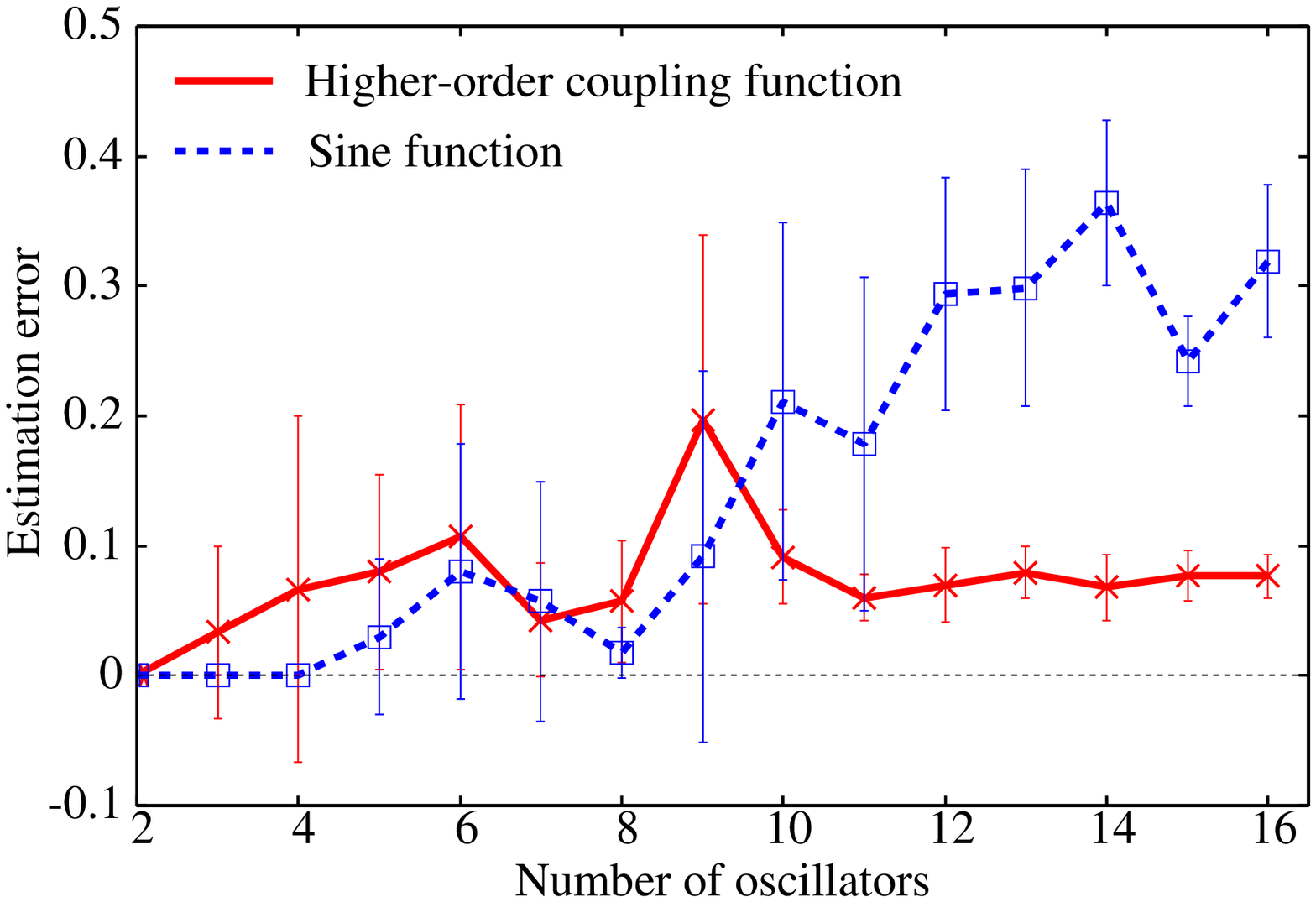}
\vspace{5mm}\\
{\scriptsize (c)}\hspace{2.5mm}
  \includegraphics[width=60mm]{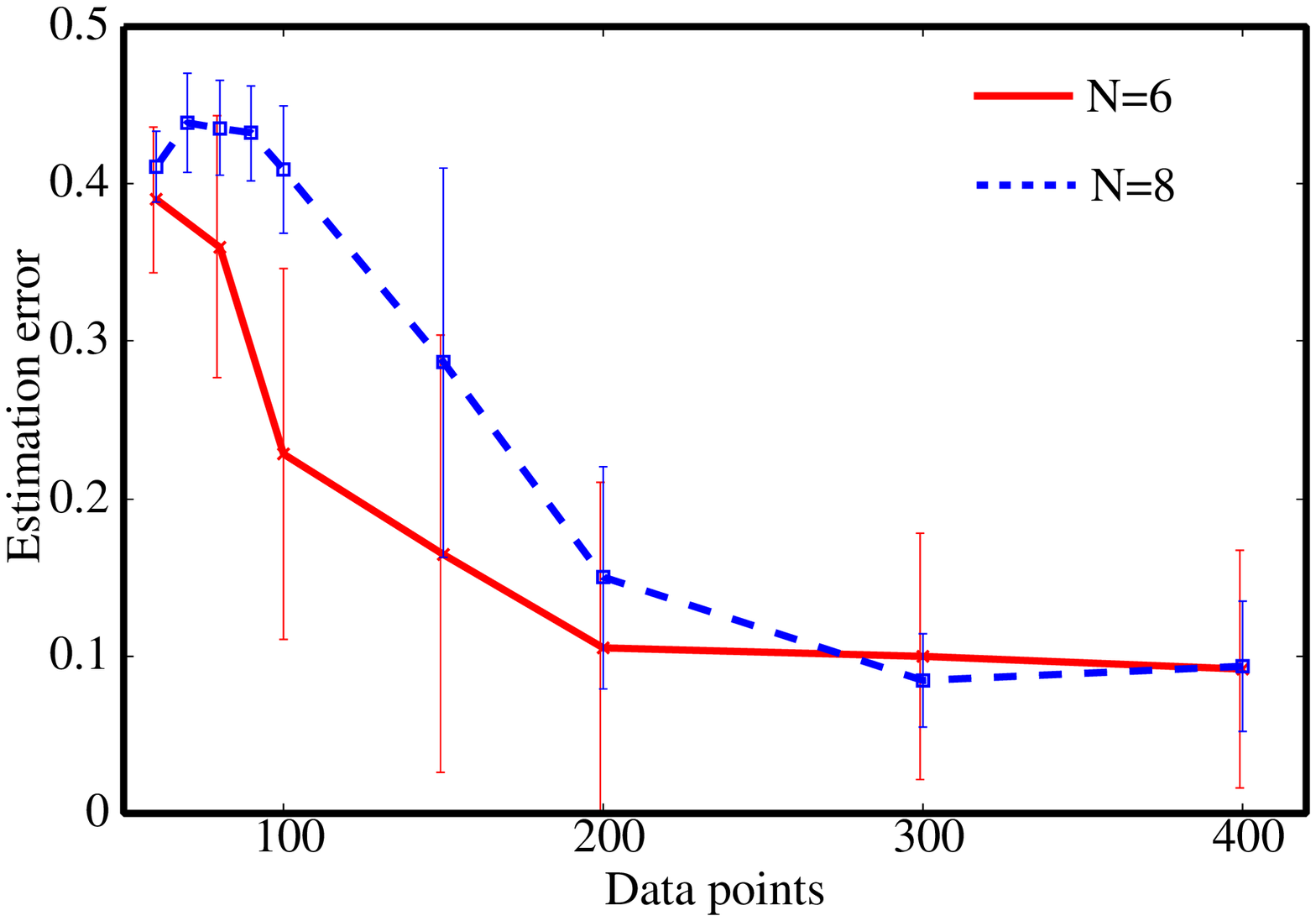}
{\scriptsize (d)}\hspace{2.5mm}
  \includegraphics[width=60mm]{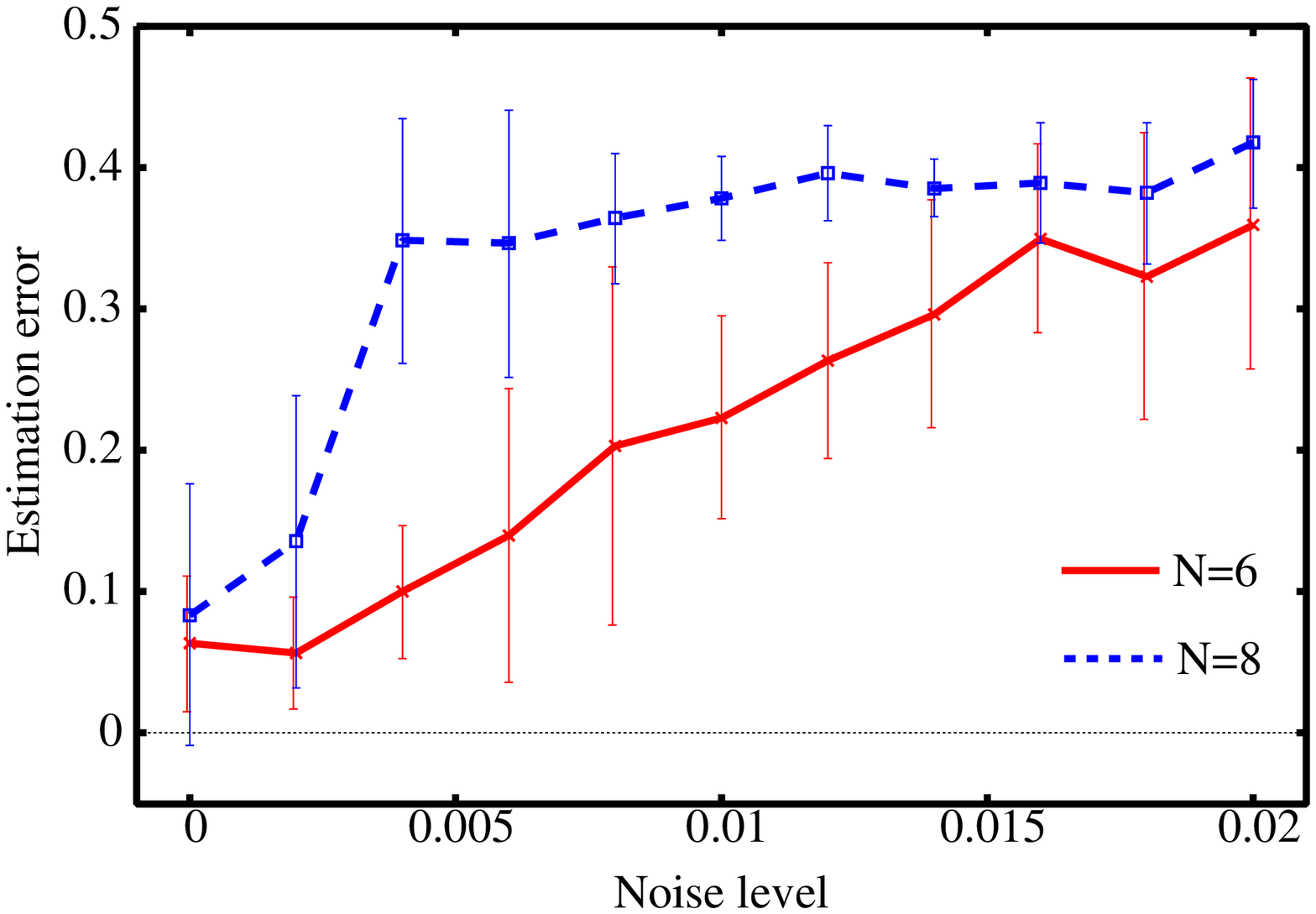}
  \caption{
Estimation errors of the network connectivity.
(a) Percentage of non-connected pairs of oscillators is $20$ \%.
The coupling function is composed of higher-order ($D=5$) 
\textit{Fourier} components in solid red line, while it is 
based on a simple sine function in dashed blue line.
(b) Percentage of non-connected pairs of oscillators is $40$ \%.
(c) Dependence of the estimation error on data length. 
Percentage of non-connected pairs of oscillators is $40$ \%,
while number of the oscillators is set to $N=6$ (solid red line)
and $N=8$ (dashed blue line).
(d) Dependence of the estimation error on noise level ${\sigma}$,
where \textit{Gaussian} noise $N(0,(2{\pi}{\sigma})^2)$ 
is added to the phase data. 
Percentage of the non-connected pairs of oscillators is $40$ \%,
while number of the oscillators is set to $N=6$ (solid red line)
and $N=8$ (dashed blue line).}
\label{fig:Networkinference}
\end{center}
\end{figure}

For comprehensive analysis, the connection matrices
with randomly generated defects were estimated for 
variable network size from $N=2$ to $N=16$. 
For our analysis, the estimation error was evaluated as
$e_{cm} = \frac{1}{N(N-1)} 
\sum_{i=1}^{N} \sum_{j=1,j{\neq}i}^{N} 
{\vert} \breve{T}_{i,j} - {T}_{i,j} {\vert}$, 
where the estimated connectivity was digitized as
$\breve{T}_{i,j}=0$ for $\tilde{T}_{i,j}<0.5$ and
$\breve{T}_{i,j}=1$ otherwise.
For each setting of the network size, $5$ instances of 
connection matrices
${\{} T_{i,j} {\}}$ were randomly generated and the 
average and the standard deviation of the estimation 
errors were plotted in Fig.~\ref{fig:Networkinference}.
Panels~(a) and (b) correspond to the cases that defect 
ratios (\textit{i.e.}, percentage of zero elements in the 
connection matrix) are $20$ \% and $40$ \%, respectively.
For variable network sizes, the estimation errors $e_{cm}$ 
are almost zero except $N=11,12$ in the case of low defect ratio. 
Although the errors increase for high defect ratio, their
overall level is less than 0.2.

To examine the effect of coupling function, the connection 
matrices were also estimated by using a simple sine as 
the coupling function, \textit{i.e.},
$\tilde{H} ({\Delta}{\theta})$ $=$ 
$a_1 {\sin} {\Delta}{\theta}$. 
For small networks, the difference was not large between
the precisely estimated (higher-order) coupling function 
(red solid line) and the simple sine function (blue dotted line).
However, as the network size increases, the estimation
error increases much more rapidly in the sine function 
than in the higher-order coupling function. 
This indicates that, for reliable detection of the 
connectivities, precisely estimated coupling function is 
of significant importance.

In Fig.~\ref{fig:Networkinference}~(c), dependence of 
the network inference on data length $M$ is indicated. 
For network sizes of $N=6$ and $N=8$, we have varied the data 
length and studied how it affected the estimation results 
of the network connectivity.
The defect ratio was set to $40$ \%.
The network inference was reliable for data length longer
than 200 points (\textit{i.e.}, about 20 cycles). 
For shorter data length, the estimation error gradually
increased. It is therefore crucial to utilize enough data 
length for precisely detecting the network connectivity.

Fig.~\ref{fig:Networkinference}~(d) shows 
dependence of the network inference on \textit{Gaussian} noise 
$N(0,(2{\pi}{\sigma})^2)$ added to the phase data. 
The defect ratio and the data length were set to $40$ \% 
and $M=400$, respectively. 
The estimation error increased gradually as the noise level 
was increased, where ${\sigma}=0.5$ \% and ${\sigma}=2$ \% 
of phase noises caused 
severe damages on the network inference for system sizes of 
$N=8$ and $N=6$, respectively. 
This suggests that our estimation technique is rather sensitive
to the phase noise and, for reliable estimation of the 
connection matrix, phase information should be accurately
extracted from the observed time series.

\section{Further Applications}
In this section, we discuss 
further applications of the multiple-shooting technique.

\subsection{Inferring Phase Sensitivity Function}
\begin{figure}[tb]
\begin{center}
{\scriptsize (a)}\hspace{2.5mm}
  \includegraphics[width=60mm]{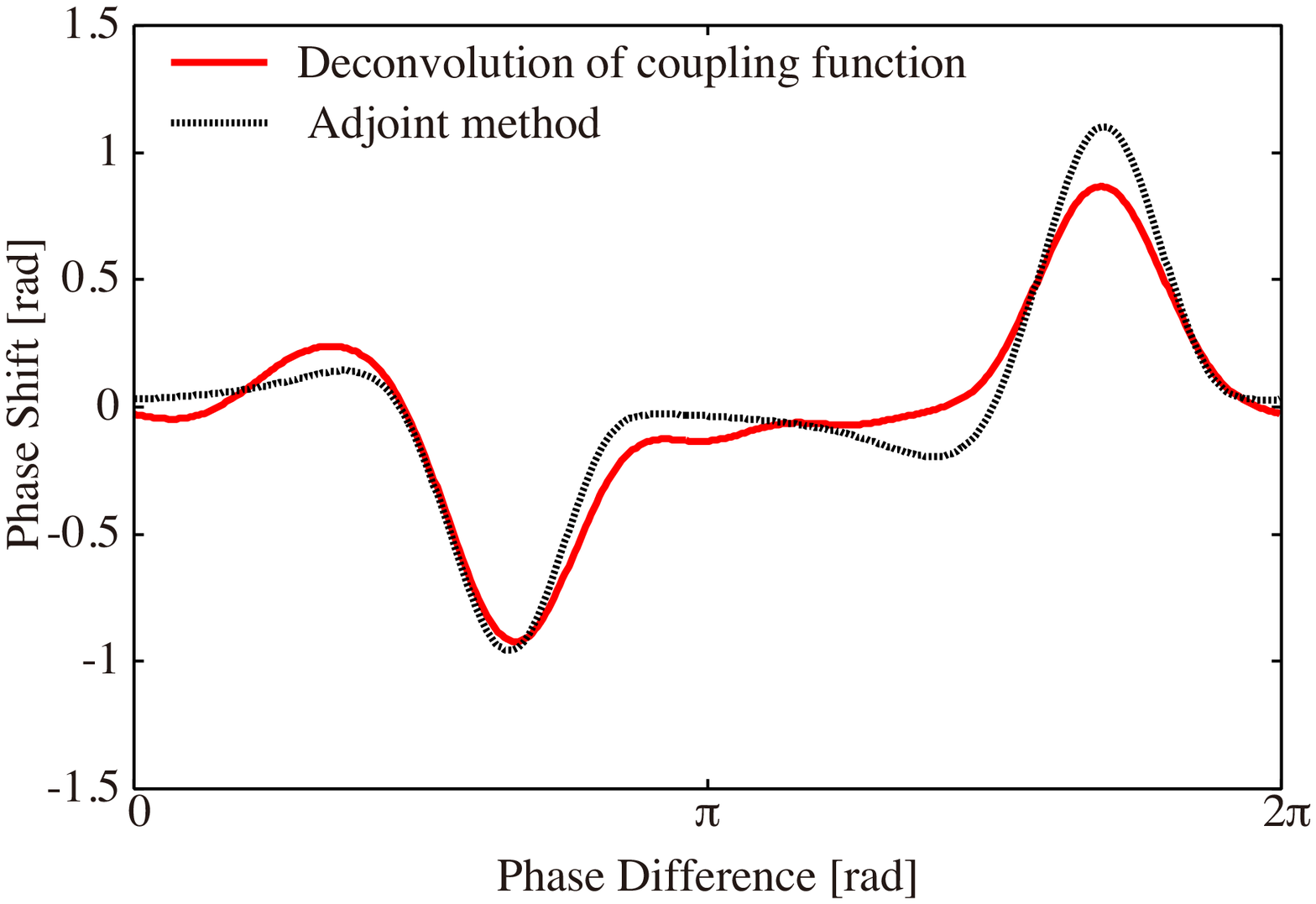}
{\scriptsize (b)}\hspace{2.5mm}
  \includegraphics[width=60mm]{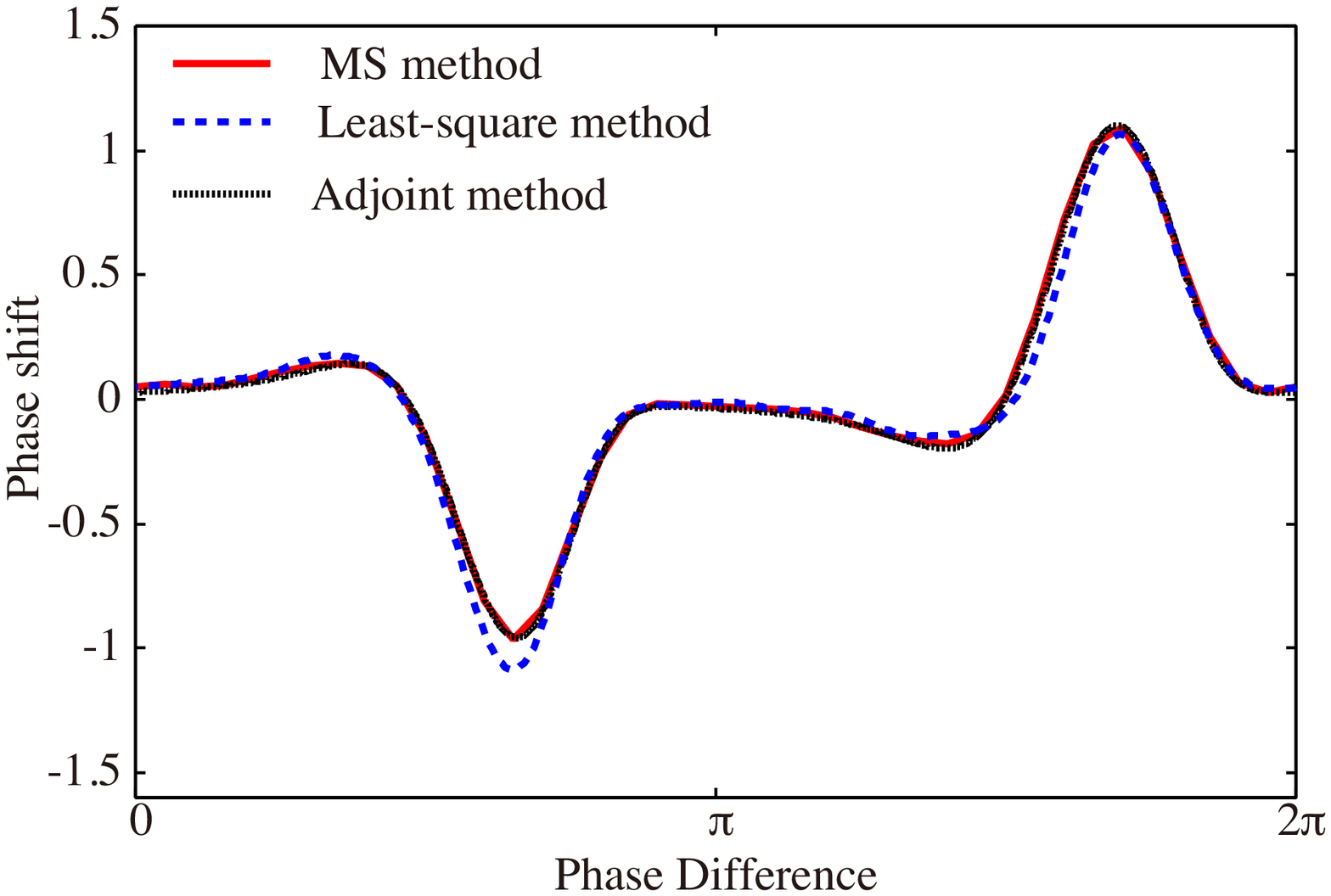}
\vspace{5mm}\\
{\scriptsize (c)}\hspace{2.5mm}
  \includegraphics[width=60mm]{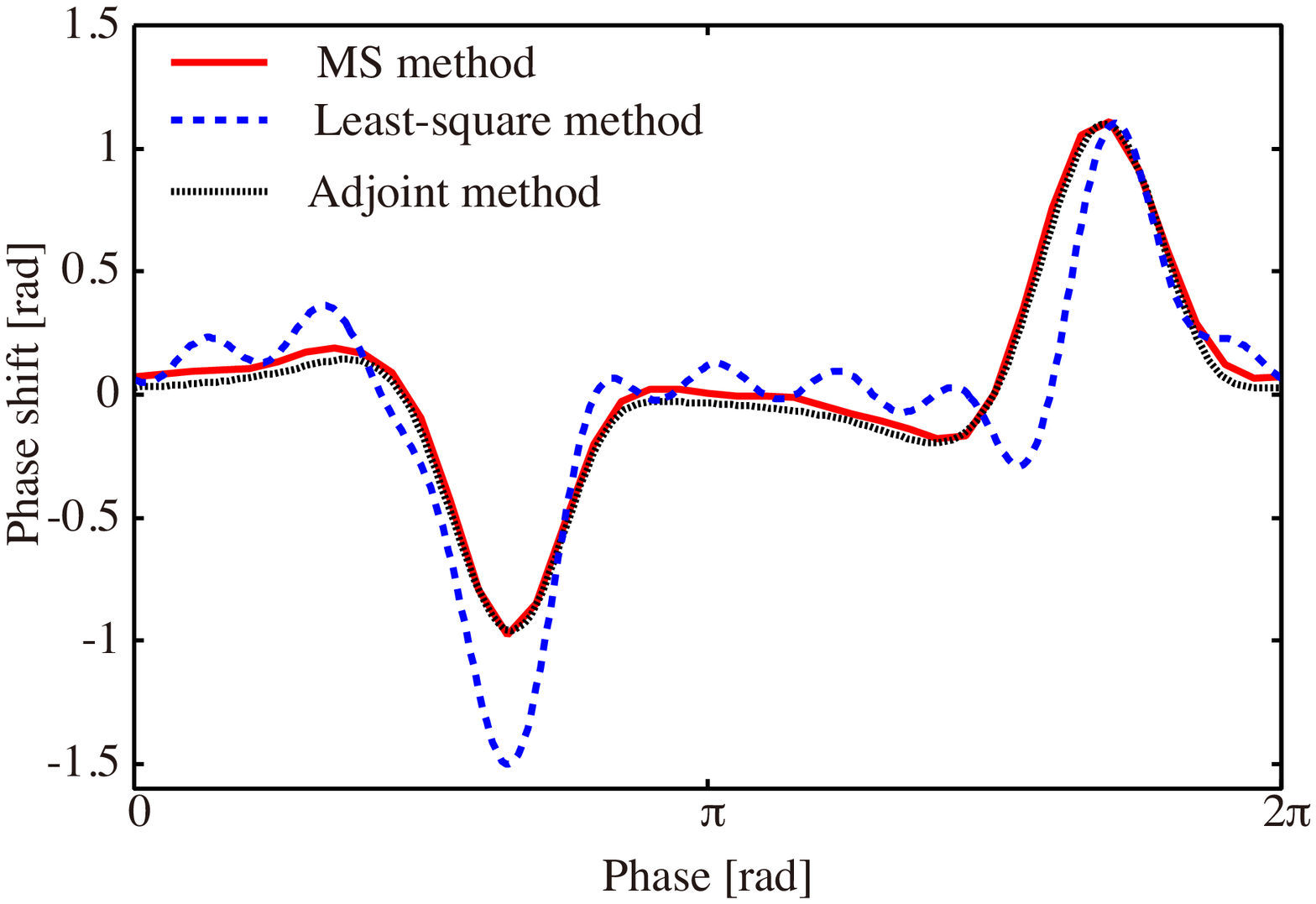}
{\scriptsize (d)}\hspace{2.5mm}
  \includegraphics[width=60mm]{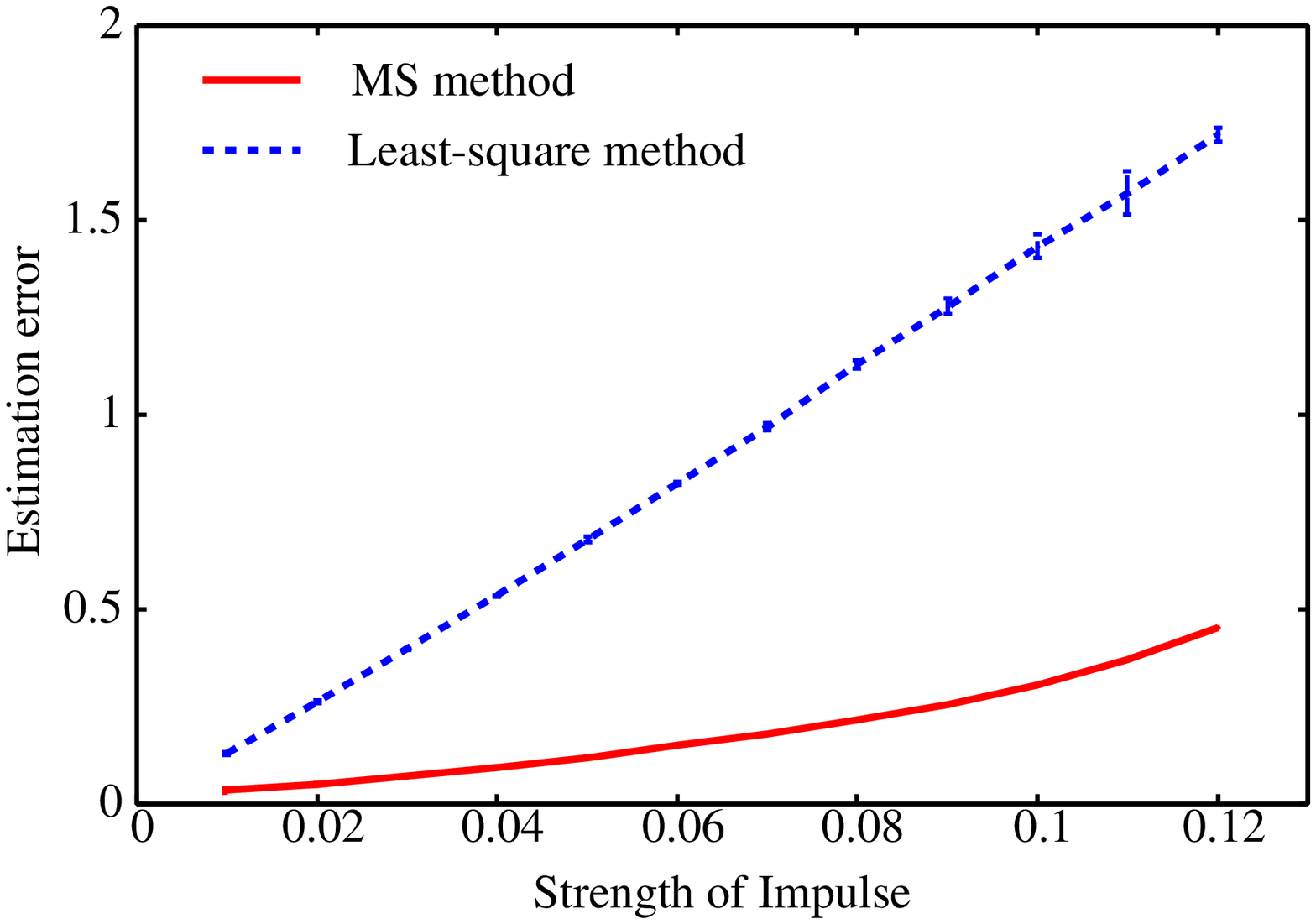}
  \caption{
(a) Phase sensitivity function $Z$ (solid red line) 
obtained by deconvolution of the coupling function 
estimated in Fig.~1a. 
Compared is the estimate by the adjoint method 
(dotted black line).
(b,c) Phase sensitivity functions $Z$ obtained by 
MS method (solid red) and the least-square method 
(dashed blue line).
Strength of the impulse is $E=0.01$ in (b) and 
$E=0.04$ in (c).
(d) Dependence of the estimation errors $e$ of
MS method (solid red line) and least-square method 
(dashed blue line) on strength $E$ of the impulses 
injected to the FHN oscillator.}
\label{fig:PRC}
\end{center}
\end{figure}

First, we apply the multiple-shooting method to the estimation 
of phase sensitivity function $Z({\theta})$.
The phase sensitivity function $Z({\theta})$ plays 
a vital role in the studies of coupled oscillators,
since it describes one of the most fundamental properties 
of the oscillator element
\cite{winfree2001,kuramoto1984,pikovsky2003synchronization}.
Numerous approaches have been proposed to estimate the
phase sensitivity function from experimental data
\cite{galan2005efficient,ota2009map,nakae2010bayesian,hong2012efficient,goldberg2013spectral,saifee2015estimation,imai2017robust,cestnik2018inferring,okada2019acoustic,nakae2019statistical}.
As an extension of our technique, the phase sensitivity 
function can be recovered from the coupling function
\cite{kralemann2013detecting}.
As described earlier in the averaging approximation 
\cite{kuramoto1984}, 
the coupling function is given by a convolution of 
the phase sensitivity function $Z({\theta})$ 
and the input waveform $G({\theta})$ as 
$H({\theta})$ $=$ $\frac{1}{2{\pi}}$ 
${\int}_{0}^{2{\pi}} Z({\psi}) G({\theta}+{\psi}) d{\psi}$
$=$ $(Z{\ast}G)({\theta})$.
It is straightforward to recover the phase sensitivity 
function by the spectral deconvolution 
\cite{wiener1949extrapolation}.
Namely, in a frequency domain, 
the phase sensitivity function is given as
$\hat{Z}({\omega}) = \hat{H}({\omega})/\hat{G}({\omega})$,
where $\hat{Z}({\omega})$, $\hat{H}({\omega})$, and
$\hat{G}({\omega})$ represent \textit{Fourier} transforms 
of $Z$, $H$, and $G$, respectively. 
Inverse \textit{Fourier} transform of $\hat{Z}({\omega})$
yields the phase sensitivity $Z({\theta})$.
Fig.~\ref{fig:PRC}(a) shows phase sensitivity function 
(solid red line) obtained by the deconvolution of the 
coupling function estimated from coupled FHN oscillators 
($N=16$) in section~2. 
Compared with the one computed by the adjoint method 
\cite{ermentrout1996type}
(dashed blue line), the estimated function is somewhat 
deviated from the true curve.
We consider that, due to the averaging effect, where
the effect of input signal is averaged over one 
oscillation cycle, information on the spontaneous phase 
response has been lost.

To improve the situation, the phase sensitivity can be
estimated more directly by using the Winfree formula
\cite{winfree2001} as follows.
For simplicity, we consider a single phase oscillator
receiving $l$-th external force $G_l(t)$
($l=1,2,{\ldots},L$):
\begin{eqnarray}
\dot{\theta}_{l} &=& {\omega} + 
\tilde{Z} ({\theta}_{l}) G_{l}(t),
\label{eq:forcedmodel}
\end{eqnarray}
where ${\theta}_{l}$ and ${\omega}$ represent phase and
natural frequency of the oscillator.
Without loss of generality, the initial phase can be
set to zero (\textit{i.e.}, ${\theta}_{l}(0)=0$). 
The external force $G_l(t)$ is typically composed of 
a short pulse, which lasts within one oscillator cycle 
of $T=2{\pi}/{\omega}$. 
The phase sensitivity function $\tilde{Z}$ is described 
in terms of a \textit{Fourier} series as
$\tilde{Z} ({\theta})$ $=$ $c_{0}$ $+$
$\sum_{j=1}^{D} c_{j} {\sin} j {\theta} + 
d_{j} {\cos} j {\theta}$.
The unknown coefficients 
$\bm{p} = {\{} c_j, d_j {\}}$
can be estimated by the multiple-shooting method in a 
similar manner as the estimation of coupling function. 
Given the external force $G_{l}(t)$, 
the phase oscillator model~(\ref{eq:forcedmodel}) can be
integrated as ${\phi}^{T} ({\theta}_{l}(0), G_{l}, \bm{p})$.
The parameters $\bm{p}$ can be optimized in such a way that
the phase model~(\ref{eq:forcedmodel}) satisfies the 
boundary conditions: ${\theta}_{l}(T)$ $=$
${\phi}^{T} ({\theta}_{l}(0), G_{l}, \bm{p})$,
where ${\theta}_{l}(T)$ represents the oscillator phase 
observed at $t=T$.

Below, we compare the performance of multiple-shooting 
method with that of least squares as the standard method 
of estimating the phase sensitivity function 
\cite{galan2005efficient,hong2012efficient}.
Here, the phase model~(\ref{eq:forcedmodel}) is integrated as
\begin{eqnarray}
 {\int}_{0}^{T} d{\theta}_{l} &=& {\int}_{0}^{T} {\omega} dt +
 {\int}_{0}^{T} \tilde{Z}({\theta}_{l}) G_{l}(t) dt ,
\nonumber\\
 {\theta}_{l}(T) - {\theta}_{l}(0) - 2{\pi} &=& 
 {\int}_{0}^{T} \tilde{Z}({\omega}t) G_{l}(t) dt ,
\nonumber
\end{eqnarray}
where the oscillator phase is approximated as 
${\theta}_{l}(t){\approx}{\omega}t$ under the assumption
that the external force $G_{l}(t)$ is weak in
equation~(\ref{eq:forcedmodel}).
By expanding the external force into \textit{Fourier} 
series as $G_{l} (t)$ $=$ $g_{l,0}$ $+$
$\sum_{j=1}^{D} g_{l,j} {\sin} j {\omega} t + 
h_{l,j} {\cos} j {\omega} t$, we obtain 
\begin{eqnarray}
  M \mathbf p = \mathbf D, \nonumber
\end{eqnarray}
where 
\begin{eqnarray}
  M=\begin{bmatrix}
      g_{1,0}/2 & g_{1,1} & h_{1,1} & g_{1,2} & h_{1,2} & \cdots & g_{1,D} & h_{1,D} \\
      g_{2,0}/2 & g_{2,1} & h_{2,1} & g_{2,2} & h_{2,2} & \cdots & g_{2,D} & h_{2,D} \\
      \vdots & \vdots & \vdots & \vdots & \vdots & \vdots & \vdots \\
      g_{L,0}/2 & g_{L,1} & h_{L,1} & g_{L,2} & h_{L,2} & \cdots & g_{L,D} & h_{L,D} 
    \end{bmatrix}, 
  \mathbf p = 
    \begin{bmatrix}
      c_0 \\ c_1 \\ d_1 \\ \vdots \\ c_D \\ d_D 
    \end{bmatrix},\ \ \ 
  \mathbf D = 
    \begin{bmatrix}
      {\Delta}{\theta}_1 \\ {\Delta}{\theta}_2 \\ \vdots \\ {\Delta}{\theta}_L 
    \end{bmatrix} . \nonumber
\end{eqnarray}
${\Delta}{\theta}_l={\theta}_{l}(T) - {\theta}_{l}(0)$ 
represents phase shift induced by the $l$-th external force $G_l(t)$.
The unknown coefficients $\bm{p}$ can be obtained as
$\bm{p}=D M^{-1}$.

We apply the two methods to a single FHN oscillator
that receives $400$ random impulses 
(stimulus duration: ${\tau}=20$, 
 stimulus strength: $V=0.01,0.02,..,0.12$)
as external forcing $G(t)$.
Parameter values of the FHN oscillator and the 
sampling time interval were set to be the same
as those in the previous sections. 
For simplicity, natural frequency ${\omega}$ and 
the external signal $G(t)$ were assumed to be known.
Number of the \textit{Fourier} components was set to $D=10$.
The integration time was set to $T=150$.
For impulse strength of $E=0.01$ and $E=0.04$,
the estimated phase sensitivity functions are drawn in 
Fig.~\ref{fig:PRC}(b) and (c), respectively.
In both panels (b) and (c), estimation results of
the multiple shooting method (solid red lines) are 
consistent with those of the adjoint method
\cite{ermentrout1996type}.
The least-square method (dashed blue line), on the 
other hand, recovered
the phase sensitivity function faithfully for a small
impulse strength in (b). The estimate is, however, 
deviated from the other two curves for a large impulse 
strength in (c).
In fact, as the impulse strength is increased,
the estimation error increases much more rapidly in 
the least-square method (dashed blue line) 
than the multiple shooting method (solid red line)
(see Fig.~\ref{fig:PRC}d).
The least-square method 
\cite{galan2005efficient,hong2012efficient}
assumes that phase of the oscillator evolves linearly 
in time according to the natural frequency.
This approximation is effective as far as the external 
force is weak. 
If stronger perturbations are applied, inducing non-small 
phase shifts, this approximation increases the estimation 
error.
The multiple-shooting method, on the other hand, takes 
into account the phase shift induced by the external 
perturbations by faithfully integrating the phase 
equation~(\ref{eq:forcedmodel}). 
The estimation error has been therefore reduced by the
multiple-shooting method.

\subsection{Chaotic Phase Synchronization}
Next we show how the estimated coupling function can be utilized 
for modeling chaotic phase synchronization
\cite{rosenblum1996phase}.
It has been known that phases of chaotic oscillators can 
be synchronized with each other, while their amplitudes
remain irregular and uncorrelated.
Especially for phase-coherent chaos, in which rotation
center can be well-defined, the phase dynamics can be
approximated as 
$\dot{\theta} = {\omega} + {\Gamma}(A)$,
where ${\Gamma}(A)$ represents frequency modulation,
which depends upon oscillation amplitude $A$
\cite{rosenblum1996phase}.
For chaotic amplitude $A$, the term ${\Gamma}(A)$ can
be regarded as an effective noise.
In many phase-coherent systems such as the R\"{o}ssler 
equations \cite{rossler1976equation},
amplitude-dependent frequency modulation is very small, 
so the noise term ${\Gamma}(A)$ is negligible.
Phase dynamics of such chaotic attractor becomes very 
similar to those of limit cycle oscillators. 

\begin{figure}[tb]
\begin{center}
{\scriptsize (a)}\hspace{2.5mm}
  \includegraphics[width=60mm]{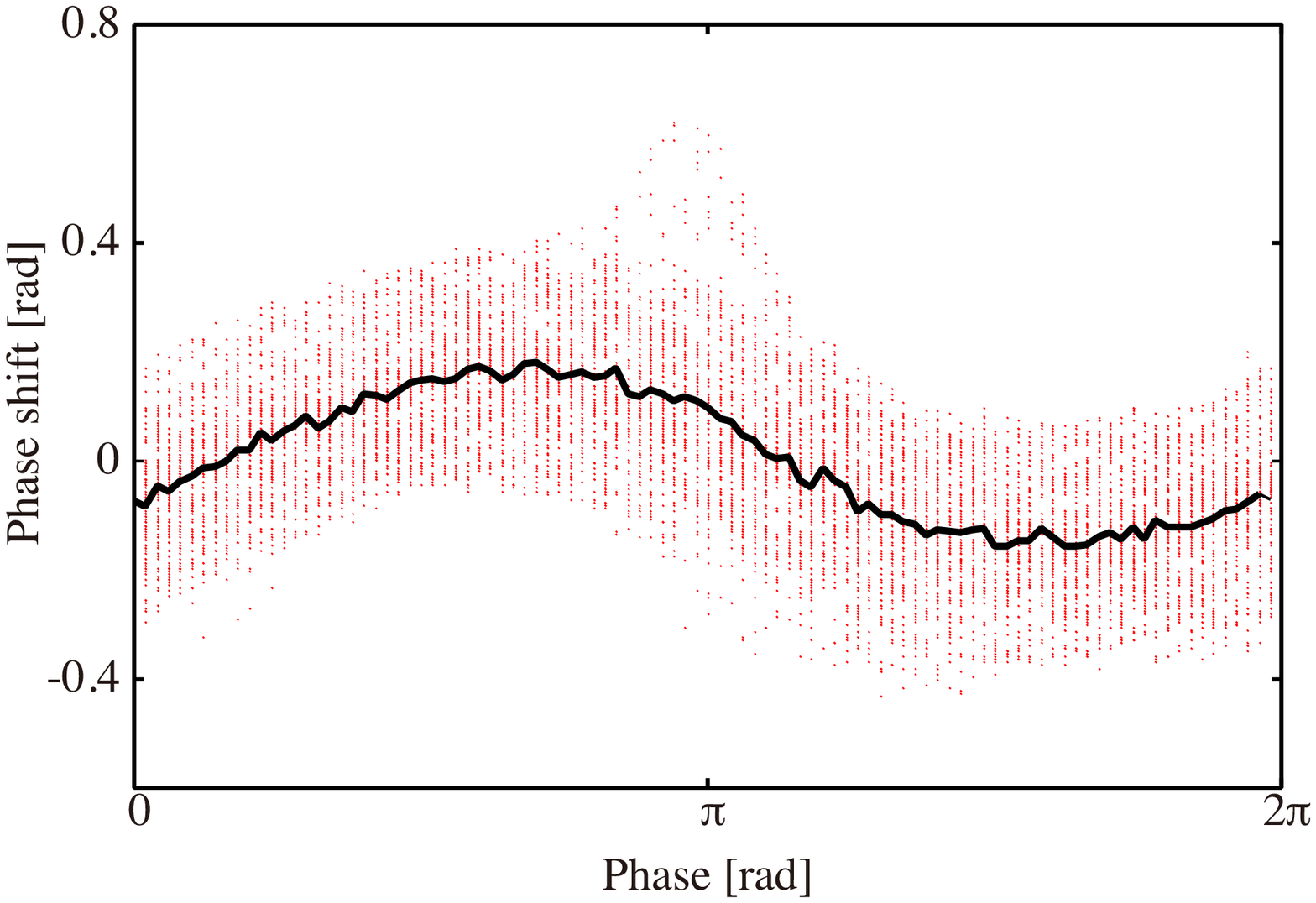}
{\scriptsize (b)}\hspace{2.5mm}
  \includegraphics[width=60mm]{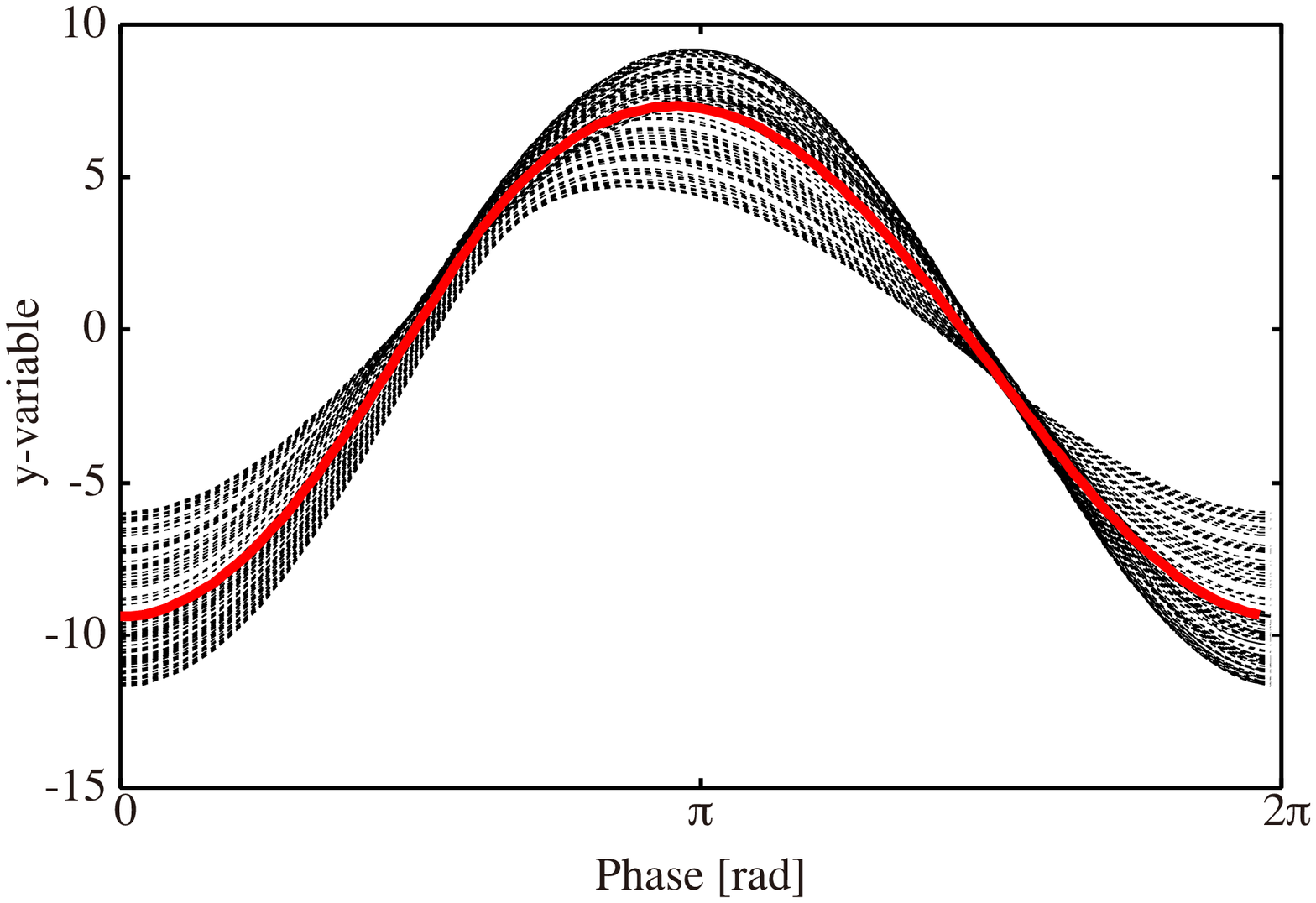}
\vspace{5mm}\\
{\scriptsize (c)}\hspace{2.5mm}
  \includegraphics[width=60mm]{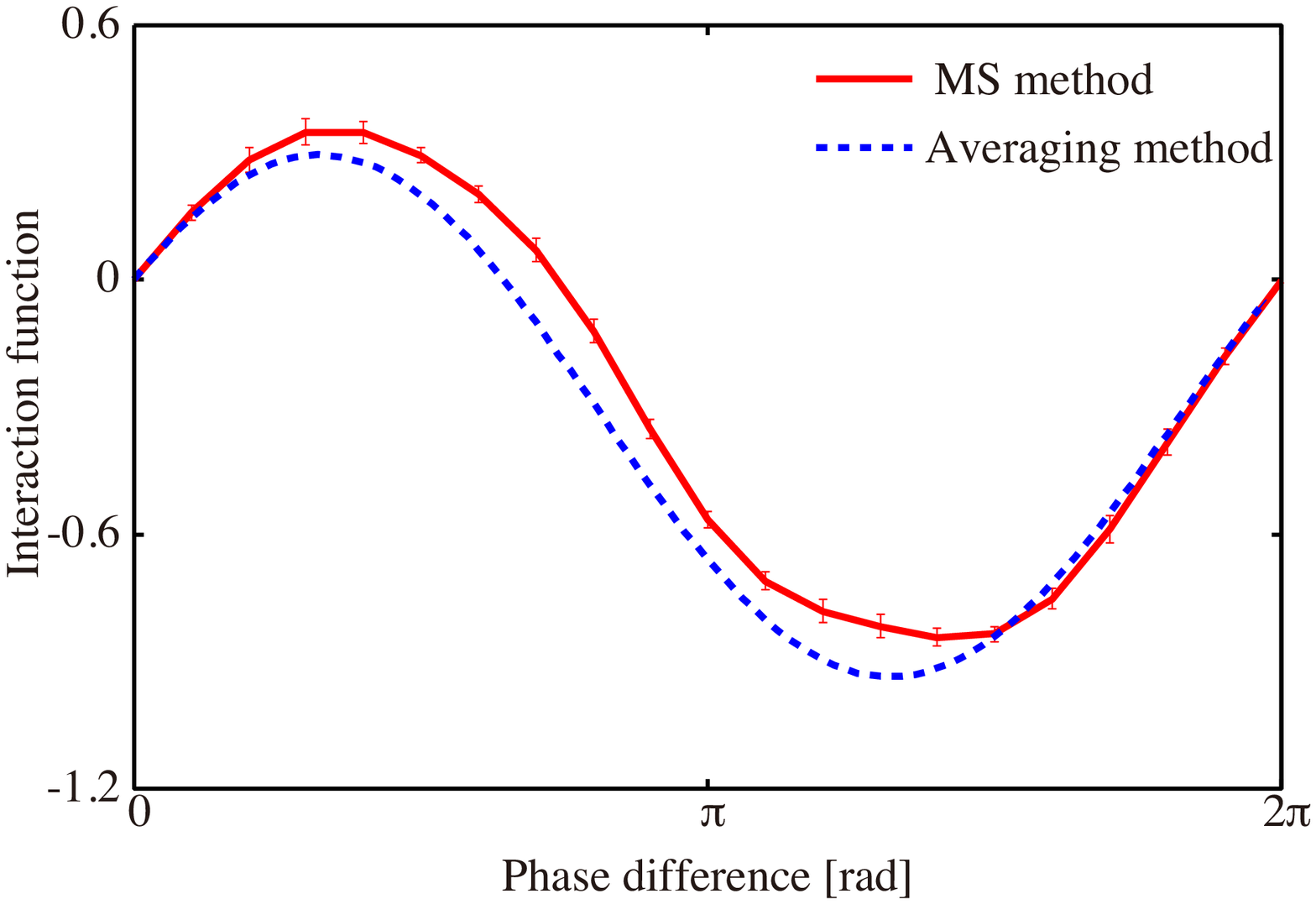}
{\scriptsize (d)}\hspace{2.5mm}
  \includegraphics[width=60mm]{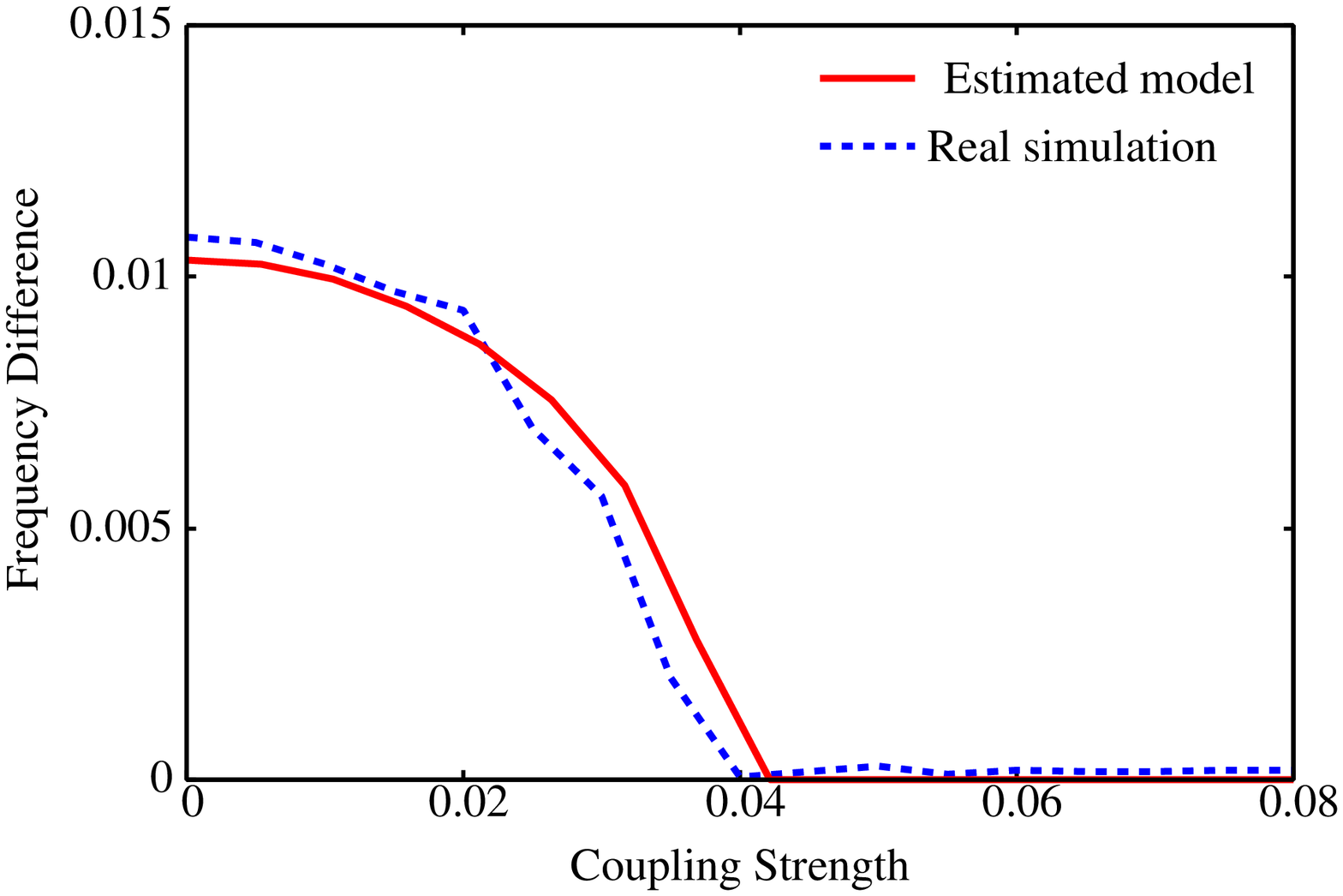}
  \caption{
(a): Phase responses of chaotic dynamics observed from 
R\"{o}ssler equations. By applying an impulse at variable
phases, the phase shifts were measured as the difference
in timing between the following peak of $y$-variable 
and the one expected from the average oscillation period.
Bold black line represents the averaged phase response.
(b): Waveforms of $y$-component of the R\"{o}ssler equations. 
Bold red line represents the averaged waveform.
(c) Coupling functions $\tilde{H} ({\Delta}{\theta})$
estimated by the present method (solid red line)
and one (dashed blue line) obtained as the convolution of 
averaged phase response curve and the averaged waveform. 
(d) Synchronization diagram of the estimated phase 
model (solid red line) and the original coupled 
R\"{o}ssler equations (dashed blue line).
}
\label{fig:ChaoticPS}
\end{center}
\end{figure}

To extract phase-interaction between chaotic oscillators,
we consider two coupled R\"{o}ssler equations 
\cite{rossler1976equation}:
\begin{eqnarray}
{\dot{x}}_{1,2} &=& -{\alpha}_{1,2} y_{1,2} -z_{1,2},
\nonumber\\
{\dot{y}}_{1,2} &=& {\alpha}_{1,2} x_{1,2}+ 0.15 y_{1,2} + 
C(y_{2,1} - y_{1,2}),
\nonumber\\
{\dot{z}}_{1,2} &=&  0.2 + z_{1,2} ( x_{1,2} - 7 ).
\nonumber
\end{eqnarray}
Each R\"{o}ssler oscillator gives rise to chaotic
dynamics without coupling $C=0$. 
The inhomogeneity parameters were set as
${\alpha}_{1,2}=1{\mp}0.01$, which yield 
average oscillation periods of 
$6.06$ and $5.94$, respectively.
The bivariate data ${\{} y_i (t) {\}}_{i=1}^{2}$ were 
simulated under coupling strength of $C=0.02$, 
which corresponds to non-synchronized regime.
The sampling interval was set to be 
${\Delta}t = 0.08$ for the extraction of
the phases ${\{} {\theta}_{i} (t) {\}}$.
Then, to apply the multiple-shooting method,
the data have been down sampled to
${\Delta}t = 1000 {\cdot} 0.08$
and the total of $2000$ data points were collected.
The data were divided into $1000$ and $1000$ points, 
which were used for the parameter estimation 
and the cross-validation test, respectively.
By varying the number of \textit{Fourier} components
from $D=1$ to $D=5$,
the optimal value was found to be $D=4$.
The corresponding coupling function 
$\tilde{H} ({\Delta}{\theta})$
is shown by the solid red line in 
Fig.~\ref{fig:ChaoticPS}(c).
The estimated function is in good agreement with the
one obtained by the convolution of averaged phase
sensitivity function (Fig.~\ref{fig:ChaoticPS}a)
and the averaged input waveform 
(Fig.~\ref{fig:ChaoticPS}b).
Using the estimated phase equations, synchronization 
diagram of the original two coupled R\"{o}ssler 
equations can be recovered, where the onset of 
synchronization was predicted at $C=0.042$, 
which is close to the real onset of $C=0.04$
(Fig.~\ref{fig:ChaoticPS}d). 
This suggests that our simple method of estimating
the coupling function provides a good approximate of
describing the phase dynamics of phase-coherent chaotic 
oscillators. 

\subsection{Application to Circuit Experiment}
\begin{figure}[tb]
\begin{center}
{\scriptsize (a)}\hspace{2.5mm}
  \includegraphics[width=60mm]{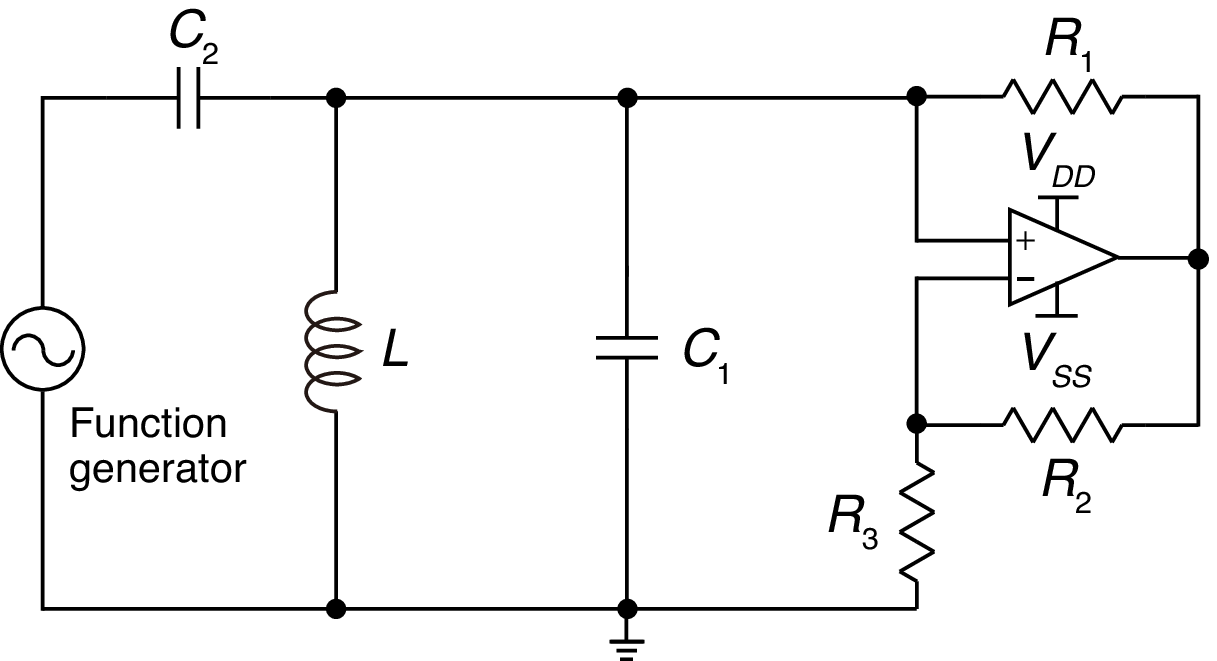}
{\scriptsize (b)}\hspace{2.5mm}
  \includegraphics[width=60mm]{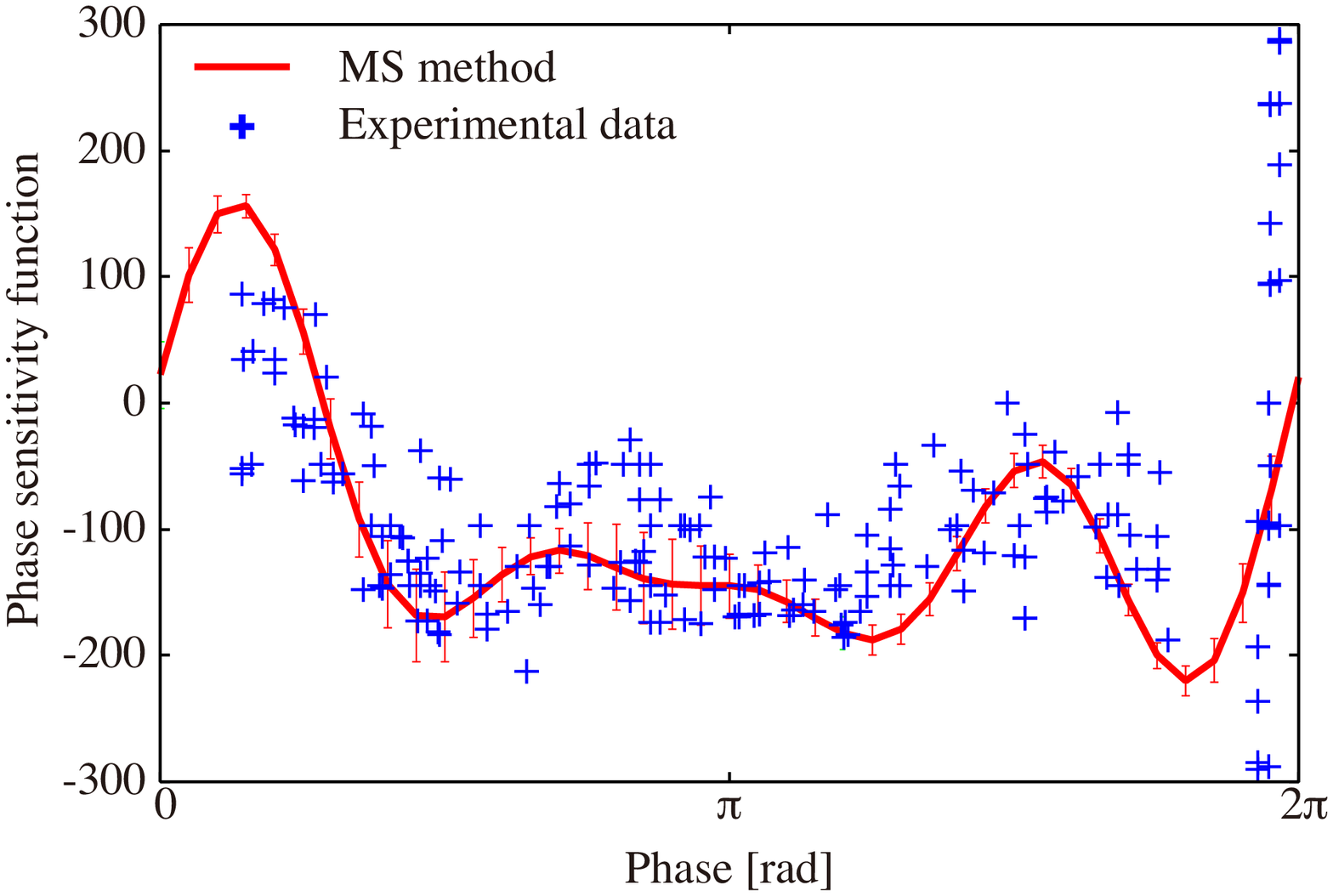}
\vspace{5mm}\\
{\scriptsize (c)}\hspace{2.5mm}
  \includegraphics[width=60mm]{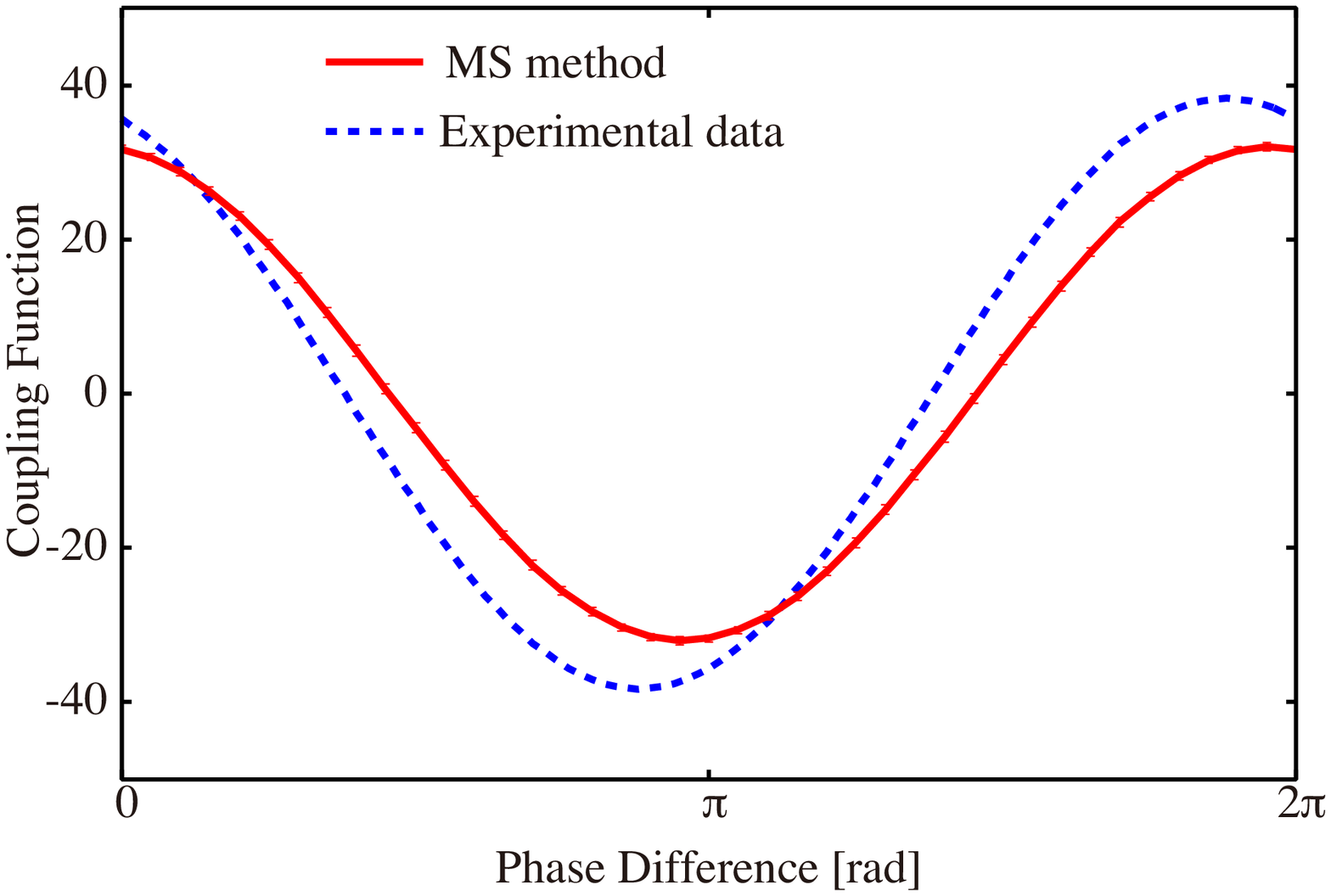}
{\scriptsize (d)}\hspace{2.5mm}
  \includegraphics[width=60mm]{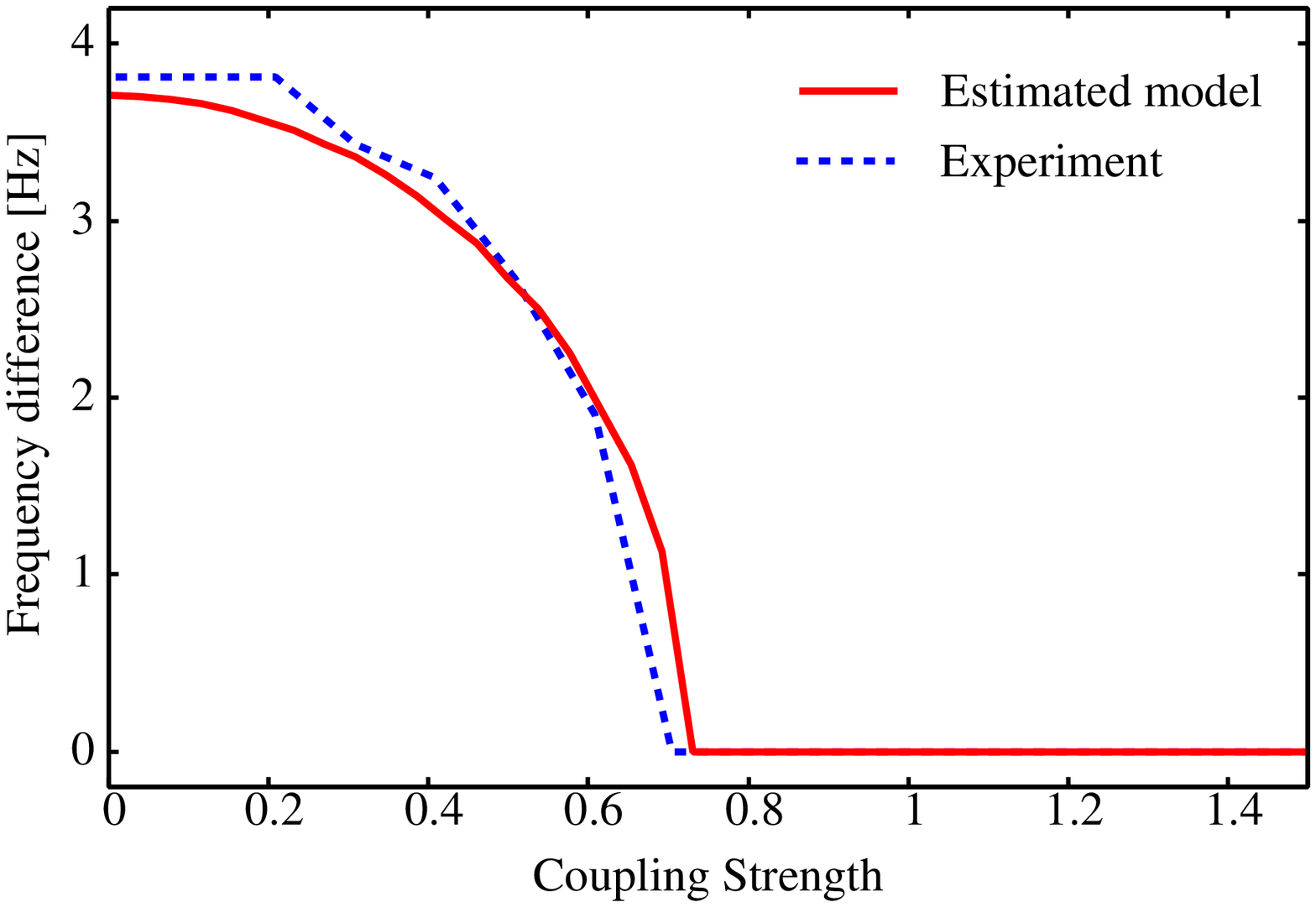}
  \caption{Experiment of Van der Pol oscillator circuit. 
(a): Schematic illustration of the Van der Pol circuit,
that is composed of an inductor ($L$), 
a capacitor ($C_1$), three resistors ($R_1$, $R_2$, $R_3$), 
an operational amplifier, and its associated
power supplies ($V_{DD}$, $V_{SS}$).
External forcing is injected from a function generator 
(Keysight 33500B) through a capacitor ($C_2$).
(b): Phase sensitivity function estimated by the 
multiple-shooting method (red line) 
and the perturbation experiment (crosses).
(c) Coupling functions $\tilde{H} ({\Delta}{\theta})$
estimated by the present method (solid red line)
and one (dashed blue line) obtained by the averaging of
the experimentally obtained phase sensitivity function 
and the sinusoidal input waveform. 
(d) Synchronization diagram of the estimated phase model 
(solid red line) and the experimental circuit system 
(dashed blue line).
}
\label{fig:Experiment}
\end{center}
\end{figure}

Finally, we apply our method to experimental data
generated from Van der Pol electric circuit 
\cite{van1927frequency}
to demonstrate the performance of our method in a realistic 
experimental setting.
As shown in the diagram of Fig.~\ref{fig:Experiment}a,
the system is based on a LC circuit, composed of an 
inductor ($L$) and a capacitor ($C_1$).
To form a negative-resistance converter, three positive 
resistors ($R_1$, $R_2$, $R_3$)
were connected to a voltage-controlled voltage source
(\textit{i.e.}, operational amplifier and its associated
power supplies $V_{DD}$, $V_{SS}$)
\cite{kennedy1992robust}.
External forcing $G(t)$ was injected from a function 
generator (Keysight 33500B) to the Van der Pol circuit 
through a capacitor ($C_2$).
Physical parameters of the electric components used in 
the present experiment are summarized in Table~\ref{tab:param}.
To obtain the phase sensitivity function, $220$ impulses 
(stimulus duration: ${\tau}=380$ $\mu$s, 
 stimulus strength: $V=3$ V)
were randomly injected as the external force $G(t)$.
The circuit output as well as the input impulses 
were simultaneously measured with a sampling
frequency of $12.5$ kHz.
First, the phase sensitivity function $\tilde{Z}$ was 
estimated by fitting the phase model~(\ref{eq:forcedmodel})
to the measured data with the multiple-shooting method. 
Natural frequency $f_{n}=110.5$ Hz
(\textit{i.e.}, ${\omega}=2{\pi}f_{n}$), 
measured before the stimulus experiment, was used in
the phase dynamics.
Number of the \textit{Fourier} components was set to $D=4$.
As shown in Fig.~\ref{fig:Experiment}b, the estimated 
phase sensitivity $\tilde{Z} ({\theta})$ fits to the 
experimental observation of phase response data well.

Next, a sinusoidal forcing 
$G(t)=V{\sin}({\Omega}t)$
(forcing frequency: $106$ Hz,
 forcing amplitude: $V=0.6$ V)
was applied to the Van der Pol circuit.
The circuit output as well as the forcing waveforms were 
simultaneously measured with a sampling
frequency of $12.5$ kHz.
By the multiple-shooting method, which fits 
the phase equations~(\ref{eq:model_phaseeq1})
to the measured data,
the coupling function $\tilde{H}$ 
(number of \textit{Fourier} components: $D=1$)
was estimated.
In Fig.~\ref{fig:Experiment}c, the estimated coupling
function is compared with the one obtained by the
averaging of the phase sensitivity function $\tilde{Z}$,
estimated from the impulse stimuli, and the input 
sine waveform $G(t)$.
Despite a slight difference in the initial phase, 
the coupling functions agree quite well with each other.
In Fig.~\ref{fig:Experiment}d, the estimated phase 
equations recovered synchronization diagram of the 
experimental system, where the onset of synchronization 
was predicted at $V=0.73$ V, which is very close to the 
real onset of $V=7$ V. 

\begin{table}[h]
\begin{center}
\caption{Parameters of Van der Pol circuit.}
\vspace{1em}
\begin{tabular}{@{\hspace{2em}}c@{\hspace{1.5em}}|l@{\hspace{2em}}}\hline
$L$ & $500$ [mH]\\
$C_1$ & $2.2$ [uF]\\
$R_1$ & $2.543$ [K$\Omega$]\\
$R_2$ & $62.7$ [k$\Omega$]\\
$R_3$ & $10$ [k$\Omega$]\\ 
$V_{DD}$ & 5 [V]\\
$V_{SS}$ & -5 [V]\\
OPAMP & LF412CN\\
$C_2$ & $10$ [nF]\\
\hline
\end{tabular}
\label{tab:param}
\end{center}
\end{table}

\section{Discussions and Conclusions}
The multiple-shooting method has been focused on as 
a non-invasive approach to estimate coupling functions 
from multivariate time series measured from a real or 
synthetic complex dynamical system
\cite{tokuda2007inferring}.
Among various methods developed so far
\cite{miyazaki2006determination,kralemann2007uncovering,kralemann2008phase,cadieu2010phase,tokuda2010predicting,kralemann2011reconstructing,blaha2011reconstruction,zhu2013quantifying,stankovski2017coupling,suzuki2018bayesian},
which are based on the Bayesian estimation and other 
variants,
the multiple-shooting provides a straightforward approach 
to fit the phase equations to phase data measured from 
an oscillator network. 
Despite its simplicity, the method was shown to be capable
of precisely estimating the coupling function of the
coupled FHN oscillators including higher-harmonic terms.
The estimation was found effective for a large network of
up to $512$ oscillators. 
Utilization of the transient part of data successfully 
enlarged applicability of the estimation technique even 
in a synchronized regime of coupled oscillators.
The estimated coupling function was further applied to 
inference of network topology and chaotic phase synchrony.
Precise estimation of the coupling functions was shown 
to improve the reconstruction of network topology.
As another intriguing issue, estimation of the phase 
sensitivity function was also discussed. 
Although the phase sensitivity function obtained by 
deconvolution of the estimated coupling function was 
slightly deviated from the true function, refinement 
has been made by extending the multiple shooting method 
directly to the phase data of a driven limit cycle 
oscillator. 
Finally, efficiency of the present approach was demonstrated
with the experimental data measured from the 
Van der Pol electric circuit with a sinusoidal forcing.
%To our knowledge, phase response property of the 
%Van der Pol circuit has not been measured experimentally,
%adding another example of PRC for experimental relaxation 
%oscillator.

Beyond experimental system in physics, chemistry, and engineering, 
we foresee that our method will be applicable to system of
rhythmic, interacting elements such as 
cellular gene expressions in the suprachiasmatic nucleus 
(SCN) \cite{yamaguchi2003}, electrical activities of cardiac 
pacemakers \cite{verheijck1998pacemaker},
inferior olive neurons in the cerebellum \cite{tokuda2017new}
and can give insights useful for domain-scientists in biology and 
neuroscience.

While considering our method of potentially practical use
for various systems, its usefulness is not without limitations. 
The main among them is the assumption that the studied system 
can be approximated as a network of weakly coupled limit cycles
\cite{kuramoto1984}. 
This, however, is not true for all systems encountered in nature. 
For instance, in gene regulatory networks, phases of the clock 
component genes are tightly connected to each other 
\cite{zhang2014circadian}.
It has been known that cortical neurons fire with a strong 
synchrony during epileptic seizure
\cite{velazquez2000gap}. 
Such strongly coupled systems should be carefully distinguished 
and avoided as a target of modeling the phase dynamics.
In the case that the system property is not well 
understood,
it is nontrivial to judge only from the recorded data
whether the coupling is weak enough to apply the phase 
modeling to the oscillator network.
It is an important open problem to provide a criterion to 
assess whether the phase model is suitable for analyzing 
the observed time series without prior knowledge on the 
underlying dynamical equations.

Another limitation is the length of the available time series: 
namely, experimental measurements, for a variety of realistic 
reasons, could produce the data (time series) of only a very 
short length. For instance, time resolved data on gene regulation 
are not likely to yield time series with much more than 
10 cycles. In this case, our method might be of limited use.
Also, realistic data are almost always noisy. The noise strength, 
depending on the experimental scenario, could be quite severe. 
Especially, the phase extraction process in our modeling is 
rather sensitive to noise. 
Temporal fluctuation and noise in natural frequencies of the
oscillator elements may also cause estimation error in 
the coupling functions.
In this respect, noise tolerance should be carefully examined, 
before the application to data contaminated with 
observational/dynamical noise.

Also, networks in real world are large and only partials of 
the dynamics elements are observable. 
Although our method was shown to be robust against system size
as far as the oscillator elements are uniformly connected and
they are desynchronized, the effect of unobserved oscillator 
states should be examined carefully.
Heterogeneity and hierarchy in the coupling functions may 
require further extension of the present approach.

Finally, we conclude the paper with a brief discussion of how 
our method's performance compares to the performance of other 
methods that reconstruct coupling functions in oscillatory 
systems. Unfortunately, such comparison is not simple to make, 
since various available methods depart from very different 
hypotheses and knowledge about the system. Stronger hypotheses 
lead to better inferences, but the information on whether the 
hypotheses are met is not always available. This  renders hard 
any independent comparison of reconstruction methods. One could 
argue that methods aimed at only network topology are more 
useful and precise, but such methods neglect the entire 
dynamical nature of many real networks. On the other hand,
certain methods give excellent results, but are limited to 
dynamical systems with specific properties. In fact, our method 
belong to this category, since it assumes the limit cycle
nature of the individual units. Furthermore, methods can be 
divided into invasive ones (that interfere with system's 
ongoing dynamics) and non-invasive ones (that do not). 
Again, their real merits is hard to compare, since invasive 
methods, although often non practical, will almost always give 
better results. Therefore, we here conclude that our 
reconstruction concept, although limited by the assumption of 
limit cycles, is a promising -- and above all 
\textit{practical} -- approach implementable in real experiments.

\enlargethispage{20pt}

%\ethics{Insert ethics statement here if applicable.}

\dataccess{
Experimental data generated from electric circuit are available in Dryad 
dataset (https://doi.org/10.5061/dryad.z34tmpg80).}

\aucontribute{
ITT designed the study.
ITT performed the numerical simulations and the data analysis. 
KI carried out the circuit experiments. 
ITT and ZL wrote the manuscript.
All authors read and approved the manuscript.
}

\competing{
The author(s) declare that they have no competing interests.}

\funding{
ITT acknowledges support 
by Grant-in-Aid for Scientific Research
(No. 17H06313, No. 16H04848, No. 16K00343, No. 18H02477)
from Japan Society for the Promotion of Science (JSPS).
ZL acknowledges support 
by EU via H2020-MSCA-ITN-2015 COSMOS 642563, and 
by Slovenian research agency via P1-0383 and J5-8236.
}

%\ack{Insert acknowledgment text here.}

%\disclaimer{Insert disclaimer text here if applicable.}

%%%%%%%%%% Insert bibliography here %%%%%%%%%%%%%%

%\bibliographystyle{rsta}
%\bibliography{Reference_Coupling}

\end{document}